\documentclass[pre,preprint,showpacs,preprintnumbers]{revtex4}

\usepackage{graphicx}
\usepackage{latexsym}
\usepackage{amsmath}
\usepackage{amssymb}
\usepackage{amsfonts}
\usepackage{bm}

\newcommand{\la}{\left<}
\newcommand{\ra}{\right>}
\newcommand{\Rend}{\ensuremath{R_\text{e}}}

\newcommand{\be}{\ensuremath{b}}
\newcommand{\br}{\ensuremath{b_\text{r}}}
\newcommand{\ce}{\ensuremath{c_\text{s}}}

\newcommand{\cP}{\ensuremath{c_\text{P}}}
\newcommand{\cV}{\ensuremath{c_\text{V}}}
\newcommand{\eself}{\ensuremath{e_\text{self}}}

\newcommand{\kappaT}{\ensuremath{\kappa_\text{T}}}
\newcommand{\geff}{\ensuremath{g}}
\newcommand{\gex}{\ensuremath{g_\text{ex}}}
\newcommand{\muex}{\ensuremath{\mu_\text{ex}}}

\newcommand{\veff}{\tilde{v}}
\newcommand{\vone}{v}
\newcommand{\Gi}{\ensuremath{G_\text{z}}}

\newcommand{\overlap}{\ensuremath{\varepsilon}}
\newcommand{\overlapghost}{\ensuremath{\varepsilon_\text{sg}}}
\newcommand{\kb}{\ensuremath{k_\text{B}}}
\newcommand{\nmon}{\ensuremath{n_\text{mon}}}
\newcommand{\Nov}{\ensuremath{N_\text{ov}}}
\newcommand{\Novghost}{\ensuremath{N_\text{sg}}}

\begin{document}

\title{A finite excluded volume bond-fluctuation model:\\
Static properties of dense polymer melts revisited}

\author{J.P.~Wittmer}
\email{jwittmer@ics.u-strasbg.fr}
\affiliation{Institut Charles Sadron, 23 rue du Loess, 67037 Strasbourg C\'edex, France}
\author{A.~Cavallo}
\affiliation{Institut Charles Sadron, 23 rue du Loess, 67037 Strasbourg C\'edex, France}
\author{T.~Kreer}
\affiliation{Institut Charles Sadron, 23 rue du Loess, 67037 Strasbourg C\'edex, France}
\author{J.~Baschnagel}
\affiliation{Institut Charles Sadron, 23 rue du Loess, 67037 Strasbourg C\'edex, France}
\author{A.~Johner}
\homepage{http://www.ics-u.strasbg.fr/~etsp/welcome.php}
\affiliation{Institut Charles Sadron, 23 rue du Loess, 67037 Strasbourg C\'edex, France}

\date{\today}

\begin{abstract}
The classical bond-fluctuation model (BFM) is an efficient lattice Monte Carlo
algorithm for coarse-grained polymer chains where each monomer occupies exclusively
a certain number of lattice sites.
In this paper we propose a generalization of the BFM where we relax this constraint 
and allow the overlap of monomers subject to a finite energy penalty $\overlap$.
This is done to vary systematically the dimensionless compressibility $g$ of the solution 
in order to investigate the influence of density fluctuations in dense polymer melts on various static 
properties at constant overall monomer density.
The compressibility is obtained directly from the low-wavevector limit of the static structure factor.
We consider, e.g., the intrachain bond-bond correlation function, $P(s)$, of two bonds separated by $s$ 
monomers along the chain.
It is shown that the excluded volume interactions are never fully screened for very long chains.
If distances smaller than the thermal blob size are probed ($s \ll g$) the chains are swollen according 
to the classical Fixman expansion where, e.g., $P(s) \sim g^{-1}s^{-1/2}$. More importantly, the polymers 
behave on larger distances ($s \gg g$) like swollen chains of incompressible blobs with $P(s) \sim g^0s^{-3/2}$.
\end{abstract}

\pacs{05.10.Ln, 61.25.hk, 61.25.hp}
\maketitle

\section{Introduction}
\label{sec_intr}

\paragraph*{The bond-fluctuation model.}
The classical bond-fluctuation model (BFM) is an efficient lattice Monte Carlo (MC) algorithm 
for coarse-grained polymer chains where each monomer occupies {\em exclusively} a certain 
number of lattice sites on a simple cubic lattice \cite{BFM,DeutschDickman90,Deutsch}. 
It was proposed in 1988 by Carmesin and Kremer \cite{BFM} as an alternative to 
single-site self-avoiding walk models, which retains the computational efficiency of 
the lattice without being plagued by severe ergodicity problems. The key idea is to 
increase the size of a monomer which now occupies a whole unit cell of the lattice, 
as illustrated in Fig.~\ref{fig_BFMsketch}. The multitude of possible bond lengths and angles allows a 
better representation of the continuous-space behavior of real polymer solutions and melts. 
%
The BFM algorithm has been used for a huge range
of problems addressing the generic behavior of long polymer chains of very different molecular 
architectures and geometries: 
statics \cite{DeutschDickman90,Deutsch,Paul91a,MP94,WM94,SMK00,BFMpressure02,WMBJOMMS04,WBJSOMB07,BJSOBW07,WBM07,papRouse} and
dynamics \cite{Paul91b,WPB92,KBMB01,MWBBL03,Takayama99b} of linear chains and rings \cite{MWC96,MWB00},
polymer blends and interfaces \cite{MarcusT3D95,Blends99,Blends03},
gels and networks \cite{Sommer05},
glass transition \cite{BBP03},
(co-)polymers at surfaces \cite{CopolySurface99},
brushes \cite{WJJB94,KBWB96,WCJT96},
thin films \cite{combpoly00,MBB01,CMWJB05},
equilibrium polymers \cite{WMC98b,WBCHCKM07,BJSOBW07} and 
general self-assembly \cite{Micelles06,Cavallo08},
\ldots.
For recent reviews see Refs.~\cite{BWM04,BFMreview05}.

\paragraph*{A BFM version allowing a systematic compressibility variation.}
As sketched in Fig.~\ref{fig_BFMsketch}, we propose here a generalization of the BFM 
where we relax the no-overlap constraint and allow the overlap of monomers subject to 
a finite energy penalty $\overlap$. 
This is done to vary systematically the strength of density fluctuations in dense solutions
and melts to study their influence on static and dynamical properties.
More specifically, we want to test the standard perturbation theory 
of weakly interacting three-dimensional polymer melts \cite{DoiEdwardsBook} and to verify 
whether certain long-range correlations \cite{WMBJOMMS04,WBJSOMB07,BJSOBW07,WBM07,papRouse}, 
which have been found recently for {\em incompressible} melts, are also present in melts 
with finite compressibility. 
We will see that this is indeed the case if one considers properties on scales larger than
the screening length $\xi \approx \be g^{1/2}$ of the density fluctuations 
where $\be$ indicates the effective bond length of the chain and 
$g$ the number of monomers spanning the ``thermal blob" 
\cite{DegennesBook,DoiEdwardsBook,foot_asymlimit}. 
Interestingly, $g$ is related to the isothermal compressibility of the solution
$\kappaT$ \cite{ANS03,ANS05a,BJSOBW07}, i.e. the {\em thermodynamic} property which measures 
the strength of the density fluctuations. It can thus be determined directly experimentally 
or in a computer simulation from the low-wavevector limit of the total monomer structure factor 
\cite{McQuarrieBook,FrenkelSmitBook,foot_Gq}
\begin{equation}
G(q) 
\equiv \frac{1}{\nmon} \sum_{n,m=1}^{\nmon} \exp\left( - i \bm{q} \cdot (\bm{r}_n - \bm{r}_m) \right)  
\stackrel{q\to 0}{\Longrightarrow} g 
\equiv  T \ \kappaT \rho,
\label{eq_gdef}
\end{equation}
with $\bm{q}$ being the wavevector,
$\bm{r}_i$ the position of monomer $i$, 
$\nmon = \rho V$ the total monomer number,
$\rho$ the monomer number density,
$V = L^3$ the volume of the system
and $T \equiv 1/\beta$ the temperature.
(Boltzmann's constant \kb \ is set to unity throughout this paper.)
%
Due to the definition given in Eq.~(\ref{eq_gdef}),
$g$ is also called ``dimensionless compressibility" \cite{foot_veff}.
%
We will show that the variation of the operational parameter $x = \overlap/T$ \cite{foot_x}, 
characterizing the strength of the overlap penalty, 
allows us to scan $g(x)$ over four orders of magnitude \cite{foot_asymlimit}. 
This puts us in the position to test various theoretical predictions 
which are sketched below.

\paragraph*{Thermodynamic properties for $x \ll 1$.}
To characterize the proposed soft BFM model we will first investigate thermodynamic 
properties such as the mean overlap energy per monomer $e$, the excess chemical potential $\muex$ 
or the (already mentioned) dimensionless compressibility $g$ as functions of $x$. In the limit 
of weak overlap penalties these properties have been calculated long ago by Edwards 
theory \cite{DoiEdwardsBook} and we will compare our numerical results with his predictions. 
For later reference we postulate here the contributions to the free energy per monomer 
$f(\beta)$ relevant for this comparison
\begin{eqnarray}
\beta f(\beta)
& = & 
\beta \eself 
\nonumber \\
& + & 
\frac{1}{N} \left(\log(\rho/N)-1\right) 
\nonumber \\
& + & 
\frac{1}{2} v(x) \rho \ 
\underline{- \frac{1}{12\pi} 
\frac{1}{\xi(x)^3\rho}}
\ + \ldots.
\label{eq_fThigh}
\end{eqnarray}
%
The first term is due to the constant intrachain self-energy discussed in Sec.~\ref{sub_energy}. 
It is due to the reference energy chosen in this study and it is not accessible experimentally \cite{foot_divergence}.
%
The second and third line of Eq.~(\ref{eq_fThigh}) can be readily obtained from the literature, 
e.g., by integrating the osmotic pressure given by Eq.~(5.45) or Eq.~(5.II.5) of 
\cite{DoiEdwardsBook} with respect to the density $\rho$.
(The first line can be considered as the integration constant with respect to this integration.)
The second line represents the translational invariance of monodisperse chains of length $N$ 
(van't Hoff's law). 
Due to this contribution the compressibilities depend in general on $N$ 
as will be discussed in Sec.~\ref{sub_compress}. 
The (bare) excluded volume interaction between the monomers is accounted for by the first term
in the third line with $v(x)$ being the second virial coefficient of a solution of unconnected 
monomers.
The underlined term represents the leading correction to the previous term due to the fact
that the monomers are connected by bonds summing over the density fluctuations to quadratic order. 
As one expects \cite{DegennesBook}, the corresponding correlations of the density fluctuations
reduce the free energy by about one $\kb T$ per thermal blob of volume $\xi^3$. 
Consistently with Refs.~\cite{DoiEdwardsBook,ANS05a,papRouse} the correlation length $\xi$
of the density fluctuations has been defined here as
\begin{equation}
\xi^2(x) \equiv \be^2(x) g(x)/12 \approx \be^2(x)/(12 v(x) \rho).
\label{eq_xidef}
\end{equation}
Note that $\geff(x)$ and $\be(x)$ refer respectively to the dimensionless 
compressibility and the effective bond length of asymptotically long chains 
\cite{foot_asymlimit,foot_connectivity}.
%
%
The density fluctuation contribution to the free energy
will be demonstrated numerically from the scaling of the 
specific heat $\cV$ with respect to overlap strength $x$ and density $\rho$
(Sec.~\ref{sub_energyfluctu}).
%
We note finally that Eqs.~(\ref{eq_fThigh}) and (\ref{eq_xidef}) are supposed to apply as long as 
\begin{equation}
\Gi(x) \equiv \frac{1}{\sqrt{g(x)}\ \be^3(x)\rho} \ll 1,
\label{eq_Ginzburg}
\end{equation}
with the Ginzburg parameter $\Gi$ being the small parameter of the perturbation theory
\cite{foot_Ginzburg}. This restricts --- as we shall see --- the related predictions to 
overlap penalties with $x \ll 1$.

\paragraph*{Conformational properties of asymptotically long chains for all $x$.}
%
This paper aims ultimately to characterize an important intrachain conformational property,
the size of chain segments of arc-length $s$ in very long chains where finite-$N$ effects may be neglected
\cite{foot_fourier}. 
Specifically, we will investigate the mean-square end-to-end distance
$R^2(s) = \la \left( \bm{r}_{m=n+s} - \bm{r}_n \right)^2 \ra$
where the average is performed over all possible pairs of monomers $(n,m=n+s)$. 
The total chain end-to-end distance is $\Rend(N) \equiv R(s=N-1)$.
If appropriately plotted \cite{WBM07}, the segment size $R(s)$ will allow us to obtain 
an estimation of the effective bond length $\be(x)$ by extrapolation \cite{foot_asymlimit}.
%
%
%
Our numerical results will be compared with an analytical prediction obtained by a standard 
one-loop perturbation calculation very similar to the analytic results already
presented elsewhere \cite{WMBJOMMS04,WBJSOMB07,BJSOBW07,WBM07,papRouse}. 
Details of the calculation for general soft melts will be presented in a future paper \cite{papPs}.
Focusing in this paper mainly on computational issues, we only quote here the ``key relation" put to the test 
\begin{equation}
1- \frac{R^2(s)}{\be^2 s}  = \frac{\ce}{g^{1/2}}
\left( \frac{1}{\sqrt{u}} - \frac{\sqrt{\pi/8}}{u} 
\left( 1 -e^{2u} \mbox{erfc}(\sqrt{2u}) \right) \right),
\label{eq_Rsgen} 
\end{equation}
where we have used $u=s/g(x)$ for the reduced curvilinear arc-length and
$\mbox{erfc}(x)$ for the complementary error function \cite{abramowitz}.
The correction to Gaussianity expressed by the {\em r.h.s.} terms of the key prediction 
is always positive and corresponds to a weak {\em swelling} of the chain.
The ``swelling coefficient" $\ce = \sqrt{24/\pi^3}/\rho \be^3$ \cite{WBM07}
measures the strength of the effect. 
While Eq.~(\ref{eq_fThigh}) fails for large $x$ where the Ginzburg parameter $\Gi$
becomes of order unity, Eq.~(\ref{eq_Rsgen}) is supposed to hold for all $x$ 
provided that the considered segment length $s$ is large enough.
As explained in Sec.~II of Ref.~\cite{WBM07}, the relevant small parameter of the 
perturbation theory is here the so-called ``correlation hole potential" of the chains
\begin{equation}
u(s) \approx \Gi \sqrt{g/s} \approx 1/\sqrt{s} \ll 1 \mbox{ for } s \gg 1
\label{eq_us}
\end{equation}
rather than $\Gi$. 
For a discussion of chain end effects in incompressible melts see  Ref.~\cite{WBM07}.

\paragraph*{Check of limiting behavior.}
Although no formal derivation of Eq.~(\ref{eq_Rsgen}) is given, 
we verify briefly whether the suggested key prediction is
reasonable by discussing the limits of large and small reduced arc-length.
We remind first that according to Flory's ``ideality hypothesis" \cite{Flory49,DegennesBook}, 
polymer chains in the melt are thought to display Gaussian statistics for segment sizes somewhat larger than the persistence 
length 
\cite{DegennesBook,DoiEdwardsBook,Flory49} which implies that the {\em r.h.s.} of
Eq.~(\ref{eq_Rsgen}) is traditionally assumed to vanish (exponentially) for finite $s$. 
At variance to this, the expansion of the complementary error function \cite{abramowitz} 
for $u = s/g \gg 1$ yields
\begin{equation}
1 - \frac{R^2(s)}{\be^2 s}  \approx \frac{\ce}{\sqrt{s}}, \label{eq_Rsmelt}
\end{equation}
suggesting in fact that corrections to Gaussianity must be taken into account for all finite $s$.
As one expects, however, Gaussianity is still recovered in three dimensions if $s \to \infty$.
(Incidentally, this does not hold for effectively two-dimensional melts as may be seen from the discussion of 
ultrathin polymer films in Refs.~\cite{ANS03,CMWJB05,MKCWB07}.)
Most remarkably, the {\em explicit} compressibility dependence drops out for large $u$ for 
all $g(x)$ provided that the chains are long enough, such that $g \ll s \ll N$. 
Eq.~(\ref{eq_Rsmelt}) is precisely the relation which has been
discussed in detail both theoretically and numerically \cite{WMBJOMMS04,WBJSOMB07,BJSOBW07,WBM07} 
for highly incompressible melts with full excluded volume interactions ($x=\infty$). 
Hence, this is the expected limiting behavior if the polymer chains are renormalized
in terms of an incompressible packing of thermal blobs with $g$ monomers.
In the opposite limit of small reduced arc-lengths, $u=s/g \ll 1$, 
the expansion of Eq.~(\ref{eq_Rsgen}) yields
\begin{equation}
1 - \frac{R^2(s)}{\be^2 s} \approx \frac{\ce}{\sqrt{g}} 
\left( \sqrt{\frac{\pi}{2}} - \frac{4}{3} \sqrt{u} \right). 
\label{eq_Rsfixm}
\end{equation}
This is consistent with the classical expansion result of the chain size in terms 
of the ``Fixman parameter" 
$z \approx v \sqrt{s}/\be^3 \approx \sqrt{u}/ (\sqrt{g} \be^3\rho)$ 
\cite{DoiEdwardsBook,foot_veff}
since the deviations from ideality expressed by the last term of Eq.~(\ref{eq_Rsfixm}) 
become then proportional to $-\ce \sqrt{u/g} \approx -z$, 
in agreement with the leading correction term for the total chain size $\Rend(N)$
presented in textbooks \cite{foot_fixman}.

\paragraph*{Outline.}
In Section~\ref{sec_algo} the algorithm is introduced and some technical details
are discussed.
First we summarize the classical BFM without monomer overlap (Sec.~\ref{sub_BFMclassic}), which has
been used in previous work \cite{WBM07} providing the start configuration for the present study, 
and introduce then its generalization with finite overlap penalty (Sec.~\ref{sub_BFMsoft}).
The central Section~\ref{sec_resu} presents our numerical results starting with 
the thermodynamic properties (Secs.~\ref{sub_energy}-\ref{sub_compress}), 
in particular the dimensionless compressibility $g(x)$.
We characterize then (Secs.~\ref{sub_bonds}-\ref{sub_Ps}) various intrachain properties, 
as for instance the effective bond length $\be(x)$ or the bond-bond correlation function $P(s)$, 
as functions of $g(x)$.
A synopsis of our results is presented in Section~\ref{sec_conc}.

\section{Algorithm and technical details}
\label{sec_algo}

\subsection{Bond fluctuation model without monomer overlap}
\label{sub_BFMclassic}

\paragraph*{A lattice Monte Carlo scheme for topology conserving polymers.}
We have used the three-dimensional version of the bond fluctuation model 
\cite{Wittmann90,DeutschDickman90,Deutsch,Paul91a,WPB92} where 
each effective coarse-grained monomer is represented by
a cube of eight adjacent sites on a simple cubic lattice,
as illustrated in Fig.~\ref{fig_BFMsketch}. 
Even the partial overlap of monomers is forbidden.
The lattice constant $a$ is naturally chosen as the unit length.
Polymers of length $N$ consist of $N$ cubes connected by $N-1$ bonds, as shown in the sketch for $N=3$. 
These bonds are taken from the set 
\begin{equation}
\left\{ P(2,0,0), P(2,1,0), P(2,1,1), P(2,2,1), P(3,0,0), P(3,1,0) \right\}
\label{eq_bondset}
\end{equation}
of allowed bond vectors, where $P$ stands for all permutations and sign combinations of coordinates. 
This corresponds to 108 different bond vectors of 5 possible bond lengths 
($2$, $\sqrt{5}$, $\sqrt{6}$, $3$, $\sqrt{10}$) and $100$ possible angles between consecutive bonds.
The smallest $13$ angles do not appear for the classical BFM without monomer overlap, because excluded 
volume forbids sharp backfolding of bonds.
If only local Monte Carlo (MC) moves of the monomers to the six nearest neighbor sites are performed 
--- called ``L06" moves \cite{WBM07} --- this set of vectors ensures automatically that polymer chains cannot cross.
Topological constraints, e.g. in ring polymers \cite{MWC96}, hence are conserved and the polymer dynamics
may be expected to be of reptation type \cite{Paul91a,MWB00,KBMB01,WBM07}. It is this fact which has originally
motivated the choice of allowed bonds. We keep it for consistency with previous work although the non-crossing
constraint is irrelevant for the present study. Note that the classical BFM algorithm with L06 moves 
is strictly speaking not ergodic, since some (thermodynamically irrelevant) configurations may be easily 
constructed which are not accessible starting from an initial configuration of stretched linear chains
\cite{BWM04}.

\paragraph*{Obtaining athermal start configurations with topology violating moves.}
The algorithm is up to now athermal and the only control parameters are the monomer density $\rho$
and the chain size $N$. Melt conditions are realized for $\rho=0.5/8$, where half of the lattice
sites are occupied \cite{Paul91a}. 
We use (if not stated otherwise) periodic simulation boxes of linear dimension $L=256$ which
contain $\nmon = \rho L^3 = 2^{20} \approx 10^6$ monomers. This large system size allows to 
eliminate finite-size effects even for the longest chain lengths studied. 
Our simulations have been carried out by a mixture of local, 
slithering snake \cite{Kron65,Wall75,MWBBL03}, and double-bridging MC moves
\cite{Theodorou02,dePablo03,Auhl03,BWM04}
which allow us to equilibrate polymer melts with chain lengths up to $N=8192$. 
Instead of the more realistic but very slow L06 dynamical scheme we make so-called ``L26" jump 
attempts to the 26 sites of the cube surrounding the current monomer position. This
permits the crossing of chains which dramatically speeds up the dynamics, expecially
for long chains $(N > 512)$. 
%
Details of the equilibration procedure and possible caveats are discussed in Ref.~\cite{WBM07}. 
We stress that if these topology violating MC moves are included all possible configurations 
become accessible, i.e.~the BFM becomes fully ergodic. Concerning the {\em static} properties ergodic 
and non-ergodic BFM versions are, however, practically equivalent. 
This has been confirmed by comparing various static properties and by counting the number of monomers 
which become ``blocked" 
once one returns to the original local L06 moves.
%

\subsection{Bond fluctuation model with finite excluded volume penalty}
\label{sub_BFMsoft}

\paragraph*{The model Hamiltonian.}
Fig.~\ref{fig_BFMsketch} shows how finite energy penalties are introduced.
The overlap of two cube corners on one lattice site ($\Nov=1$) corresponds to an 
energy cost of $\overlap/8$, the full overlap of two monomers ($\Nov=8$) to an energy $\overlap$.
More generally, with $\Nov$ being the total number of interacting cube corners 
the total interaction energy of a configuration is
\begin{equation}
E = \frac{\overlap}{8} \Nov.
\label{eq_Hamiltonian}
\end{equation}
With the energies of the
final ($E_\text{f}$) and the initial configurations ($E_\text{i}$) we accept the MC move
according to the Metropolis criterion \cite{BWM04,LandauBinderBook} with probability
$\mbox{min}(1,\exp[-(E_\text{f} - E_\text{i})/T])$. 
We set arbitrarily $\overlap =1$ and vary the ratio $x = \overlap/T$ \cite{foot_x}
starting from $x = \infty$ corresponding to the athermal classical BFM and 
systematically increase the temperature $T$ 
as shown in Table~\ref{tab_beads} for unconnected monomers ($N=1$).

\paragraph*{Second virial coefficient.}
To illustrate this interaction we indicate already here the second virial of an imperfect gas of
unconnected monomers,
$\vone = \int d{\bm \delta} (1-e^{-E({\bm \delta})/T})$, 
which is shown below to be useful for roughly characterizing the effective strength of the potential. 
$\bm \delta$ stands for a possible lattice vector between the centers of two interacting cubes. 
It is easy to see that there are 8 vectors corresponding to $\Nov =1$ (as in Fig.~\ref{fig_BFMsketch}), 
12 to $\Nov=2$ (overlap of two cube corners), 
6  to $\Nov = 4$ (overlap of two faces),
and 1 to $\Nov=8$ (full overlap). 
This leads to a second virial
\begin{equation}
\vone = 8 \times  (1 - \exp(-x /8)) 
  + 12 \times (1 - \exp(-x /4))
  + 6  \times (1 - \exp(-x /2))
  + 1  \times (1 - \exp(-x ))
\label{eq_v}
\end{equation}
given in units of the lattice cube volume $a^3$.
We note that the second virial becomes constant,  
$\vone = 27$, in the low temperature limit ($x \gg 1$) 
corresponding to the classical athermal BFM result \cite{DeutschDickman90}. 
In the opposite high temperature limit ($x \ll 1$) it decays as 
\begin{equation}
\vone \approx 8 x - \frac{27}{16} x^2.
\label{eq_vThigh}
\end{equation} 

\paragraph*{Implementation of the algorithm.}
We briefly explain the implementation of BFM chains with the soft
overlap Hamiltonian, Eq.~(\ref{eq_Hamiltonian}).
Following \cite{Paul91a,Paul91b,WPB92,KBMB01,MWBBL03}
it is convenient to keep one list of the monomer positions in absolute space (since we are also
interested in dynamical properties) and one for the corresponding lattice positions.
We identify each of the 108 allowed bonds of the set [Eq.~(\ref{eq_bondset})] with a unique bond index
and keep a list of these indices. 
This allows to verify rapidly whether an attempted bond vector is acceptable.
Since the bond index (being less than 128) can be encoded as a byte, this compressed list is stored 
when we write down the configuration ensemble for further analysis. Additional lists allow to
handle efficiently the periodic boundary conditions, the change of a bond index for a given 
monomer move and the local interactions relevant for local L06 or L26 moves.

Since the soft overlap Hamiltonian allows the occupation of a lattice site by more than one
monomer, it is not possible to use a compact boolean occupation lattice
(corresponding to a spin $\sigma=0$ and $\sigma=1$ for an empty or filled lattice site) 
or an integer lattice filled with the monomer indices as in previous implementations. 
Instead we have mapped Eq.~(\ref{eq_Hamiltonian}) onto a Potts spin model \cite{LandauBinderBook} 
\begin{equation}
E = \frac{1}{2} 
\sum_{\bm{r}}
\sigma(\bm{r}) 
\sum_{\bm{\delta}} 
J(\bm{\delta}) \sigma(\bm{r}+\bm{\delta})
- \frac{1}{2} \overlap \nmon
\label{eq_spinmodel}
\end{equation}
with constant monomer number $\nmon=\sum_{\bm{r}} \sigma(\bm{r}) \stackrel{!}{=} L^3 \rho$ 
using a Wigner-Seitz representation of the cubic lattice following M\"uller 
\cite{MarcusT3D95,BFMreview05}. 
In this representation an integer spin variable $\sigma(\bm{r})$ counts the number of BFM monomers 
($\sigma=0,1,2,\ldots$) with cubes centered at a Wigner-Seitz lattice position $\bm{r}$.
In other words, each cube is not represented by 8 lattice entries for the cube corners, 
but just by one for its center. Since we have now to compute the interaction between cube centers
instead of cube corners, the coupling constant $J$ characterizing the interaction between
two spins depends only on the relative distance $\bm{\delta}$:
\begin{equation}
J(\bm{\delta}) = \overlap \left\{
\begin{array}{ll}
1/8       & \mbox{if $\bm{\delta}=P(1,1,1)$ (cube corners),} \\
1/4       & \mbox{if $\bm{\delta}=P(1,1,0)$ (cube edges),} \\
1/2       & \mbox{if $\bm{\delta}=P(1,0,0)$ (cube faces),} \\
1         & \mbox{if $\bm{\delta}=P(0,0,0)$ (full overlap),} \\
0         & \mbox{otherwise.} 
\end{array}
\right.
\label{eq_J}
\end{equation}
%
Since the interaction is still short-ranged and the values of $J$ are readily tabulated,
this remains an efficient rendering of the monomer interactions.
Note that the first term on the {\em r.h.s.} of Eq.~(\ref{eq_spinmodel}) 
contains a constant self-interaction contribution of the $\nmon$ monomers 
with themselves for $\bm{\delta}=\bm{0}$, which is substracted by the second term
\cite{foot_attraction}.

\paragraph*{Equilibration and system properties of high-molecular melts.}
As already stated we have used as start configurations the equilibrated BFM configurations 
without monomer overlap from our computationally much more expensive previous studies 
obtained with topology violating moves \cite{WBM07}. 
Since the soft BFM simulations are also ergodic, 
these are the relevant reference configurations. Starting with these configurations we 
increase the temperature.
As one may expect, the configurational properties essentially are found unchanged for 
low temperatures ($x \gg 5$).
Local L26 moves need to be added to global slithering-snakes moves for $x \ge 1$.
Otherwise the pure snake motion will become ineffective as it is well known from a previous
study of the snake dynamics without overlap \cite{MWBBL03}.
Simple slithering-snake moves without additional local moves are sufficient for smaller $x$.
We have crosschecked our results in this regime for $N=2048$ and $N=8192$ using boxes of linear size 
$L=512$  by starting our simulations with Gaussian chains at $x=0$ and increasing then the penalty.  
Tables~\ref{tab_epsasym} and \ref{tab_eps1.0} present some system properties obtained for our reference
density $\rho=0.5/8$.
%
Averages are performed over all chains and at least 100 configurations.
%
Table~\ref{tab_epsasym} summarizes the properties extrapolated for asymptotically long chains.  
Similar information is given in Table~\ref{tab_eps1.0} for a constant penalty $x=1$ 
as a function of chain length $N$.
Density effects have been studied only briefly for chains of length $N=8192$ and 
weak overlap penalties ($x \ll 1$). This has been done to investigate
the intrachain contributions to the mean energy.
We begin the discussion of our numerical results by addressing this issue.

\section{Computational results}
\label{sec_resu}

\subsection{The mean overlap energy}
\label{sub_energy}

\paragraph*{Qualitative behavior.}
From the numerical point of view the simplest thermodynamic property to be investigated here 
is the mean interaction energy per monomer, $e = \la E \ra/\nmon$, due to the Hamiltonian, 
Eq.~(\ref{eq_Hamiltonian}). Fig.~\ref{fig_Eeps} presents the dimensionless energy $y = e/\overlap$ 
as a function of the reduced overlap penalty $x=\overlap/T = \overlap \beta$ \cite{foot_x} 
for melts at monomer number density $\rho=0.5/8$ and various chain lengths $N$ as indicated.
We increase the temperature $T$ from the right to the left starting with configurations 
obtained recently \cite{WBM07} for the classical athermal BFM. As one expects, 
the interaction energy increases exponentially for small $T$ and levels off for large $T$ 
where chains and their monomers freely overlap ($x \ll 1$). 
The data for unconnected beads ($N=1$) represented by the filled spheres and polymer chains 
($N \gg 1$) are broadly speaking similar, especially for large overlap penalties, $x > 1$. 
Interestingly, the mean energy of polymer melts increases more strongly for $x \ll 1$
as can be seen better from the log-linear data representation chosen in the inset of Fig.~\ref{fig_Eeps}.
Also shown in the inset is the mean intrachain self-energy per chain monomer $\eself$ 
(filled triangles) obtained for the largest chain length $N$ available for a given $x$. 
As can be seen also from Table~\ref{tab_epsasym}, about half of the energy of polymer melts 
for all $x$ is due to these intrachain interactions.
For $x \ll 1$ the self-energy becomes $\eself/\overlap \approx 0.18$
which is exactly the observed energy difference between polymer and bead systems.

\paragraph*{Second virial contribution.}
Before addressing this point let us first consider the energy of unconnected soft BFM beads 
for which the second virial coefficient $v(x)$ has been given in Eq.~(\ref{eq_v}). 
Since $e = \partial_{\beta} (\beta f(\beta))$ the mean energy per bead becomes to
leading order \cite{McQuarrieBook} 
\begin{equation}
y \approx \frac{1}{2} \rho \frac{\partial v(x)}{\partial x} = \frac{\rho}{2} 
\left( \exp(-x/8) + 3 \exp(-x/4) + 3 \exp(-x/2) + \exp(-x) \right)
\label{eq_evirial}
\end{equation}
corresponding to the first term in the third line of Eq.~(\ref{eq_fThigh}). 
Eq.~(\ref{eq_evirial}) is represented by the dashed line in Fig.~\ref{fig_Eeps}.
It corresponds to a Arrhenius behavior with
$y \approx \rho \exp(-x/8)/2$ (dash-dotted line) in the low temperature region 
and, as expected, to $y \rightarrow \frac{1}{2} 8\rho = 4\rho$ for large temperatures.
This simple formula predicts well the bead data over the entire range of $x$
(underestimating slightly the mean energy at $x \approx 10$) and yields also
a remarkable fit for polymer chains with larger overlap penalties.

\paragraph*{Self-energy in the high temperature limit.}
The above-mentioned energy difference between polymer chains and beads for $x \ll 1$ has 
been accounted for by the first free energy contribution indicated in Eq.~(\ref{eq_fThigh}).
%
This contribution is further investigated in Fig.~\ref{fig_Erho} 
presenting data for such a high temperature ($x=0.001$) that the entropy dominates essentially 
all conformational properties. The self-energy of a chain is thus given by the probability $p(s,\bm{\delta})$ 
that a random walk of $s$ BFM bonds returns to a relative position $\bm{\delta}$ with respect 
to a reference monomer at $\bm{r}$.  The self-energy per monomer is then 
\begin{equation}
\eself = \frac{2}{N} \sum_{\bm{\delta}} \sum_{s=2}^{N-1} (N-s) J(\bm{\delta}) p(s,\bm{\delta})
\label{eq_enoninter}
\end{equation}
where the first sum runs over all positions with non-vanishing coupling constant 
$J(\bm{\delta})$ as defined in Eq.~(\ref{eq_J}). 
The probability $p(s,\bm{\delta})$ and the weights $J(\bm{\delta}) p(s,\bm{\delta})$ can be 
tabulated in principle for small $s$. Since the return probability decreases strongly with $s$, 
these  model-specific small-$s$ values dominate the integral, Eq.~(\ref{eq_enoninter}).
As can be seen from the inset of Fig.~\ref{fig_Erho} for single chains (corresponding to an
overall density $\rho=0$) the self-energy per monomer
becomes $\eself \approx 0.18 \overlap$ for large $N$.
The weak chain length dependence visible in the panel stems from the upper integration
boundary over the Gaussian return probability which leads to a chain length correction 
linear in $t \equiv 1/\sqrt{N-1}$. This is indicated by the bold line presented in the panel.
Also shown in the panel are energies for our reference density $\rho=0.5/8$.
They are shifted vertically by the mean field energy $4 \rho$ 
assuming that densities fluctuations of different chains do not couple.
The main panel presents the mean energy $e$ and the mean self-energy $\eself$
as functions of the density $\rho$ for chains of length $N=8192$. 
The self-energy (triangles) stays essentially density-independent. 
The total interaction energy sums over the self-energy and mean-field energy contributions
as shown by the dashed line. The self-energy contribution can only be neglected
for very large densities corresponding to volume fractions larger than unity.

\paragraph*{Temperature dependence in the high temperature limit.}
Summarizing Eqs.~(\ref{eq_fThigh}), (\ref{eq_xidef}) and (\ref{eq_vThigh}) the energy per bead should
scale to leading order in $x$ as 
\begin{equation}
y \approx 0.18 + 4\rho 
- \ 
\underline{ \frac{24^{3/2}}{\pi} \frac{\sqrt{x \rho}}{l^3 \rho}} 
+ \ldots
\mbox{ for } x \ll 1
\label{eq_eThigh}
\end{equation}
where the two $x$-independent contributions have already been discussed above.  
The underlined term stems from the density fluctuation contribution for long polymer chains
predicted by Edwards \cite{DoiEdwardsBook} indicated in Eq.~(\ref{eq_fThigh}). 
Here we have approximated the effective bond length $\be(x)$ by the mean-squared bond length $l \sim x^0$. 
As can be seen from Table~\ref{tab_epsasym} this approximation (further discussed in Sec.~\ref{sub_chainsize})
is justified for $x \ll 1$.
Eq.~(\ref{eq_eThigh}) is indicated by the bold line in the inset of Fig.~\ref{fig_Eeps}. 
It yields a reasonable description of the temperature dependence of the mean energy for small $x$.
Since the energy is dominated by the two constant contributions to Eq.~(\ref{eq_eThigh})
for $x \le 0.001$ and since higher expansion terms become relevant for $x  > 0.1$, 
the predicted $\sqrt{x}$-decay corresponds unfortunately only to a small $x-$regime.  
In order to show that it is indeed the density fluctuation term which dominates the 
temperature dependence for $x \ll 1$ we will consider in the next paragraph the 
specific heat $\cV$, i.e. the second derivative of the free energy with respect to $\beta$.

\subsection{Energy fluctuations}
\label{sub_energyfluctu}

\paragraph*{Specific heat.}
The fluctuations of the interaction energy are addressed in Fig.~\ref{fig_cVeps} displaying 
the enthalpic contribution to the specific heat per monomer, 
$\cV = -\beta^2 \partial^2_{\beta} (\beta f(\beta)) = 
\frac{1}{T^2} \left( \la E^2 \ra - \la E \ra^2  \right)/\nmon$
\cite{McQuarrieBook}.
Using again the second virial of soft BFM beads, Eq.~(\ref{eq_v}), one obtains 
the simple estimate for the specific heat 
\begin{equation}
\cV = \frac{\rho}{2} x^2 
\left(\frac{1}{8} \exp(-x/8) + \frac{3}{4} \exp(-x/4) + \frac{3}{2} \exp(-x/2) + \exp(-x) \right)
\label{eq_cVvirial}
\end{equation}
represented by the dashed line. 
In the large-$x$ limit, this corresponds to the exponential decay,
$\cV \approx \rho x^2 \exp(-x/8)/16$, indicated by the dash-dotted line.
For barely interacting beads ($x \ll 1$), Eq.~(\ref{eq_cVvirial}) yields a power-law limiting
behavior, $\cV \approx \frac{27}{16} \rho x^2 \sim 1/T^2$. 
As can be seen from the plot, Eq.~(\ref{eq_cVvirial}) predicts the energy fluctuations
of BFM beads for essentially all $x$, slightly underestimating again the maximum of $\cV$ at $x \approx 10$. 
Since the specific heat for beads and polymer chains is similar for $x > 1$,
the virial formula is also good for polymer chains in this limit.

\paragraph*{High temperature limit for polymer melts.}
Strong chain length effects are, however, visible for high temperatures
($x \ll 1$) where the specific heat is found to increase monotonously with $N$.
This can better be seen from the inset where the specific heat is plotted
as a function of the reduced chain length $u=N/g$ with $g$ being the dimensionless
compressibility determined below in Sec.~\ref{sub_compress}.
(Since $e$ and $\cV$ correspond to different derivatives of the free energy $f$
with respect to $\beta$, there is obviously no thermodynamic inconsistency in the finding that $\cV$ 
reveals much larger chain length effects than $e$.)
For large chains with $u \gg 1$ this increase levels off at a chain length independent envelope 
\begin{equation}
\cV \approx 
\frac{24\sqrt{6}}{\pi} \frac{\rho^{1/2}}{l^3} x^{3/2} N^0
+ \ldots 
\label{eq_cVThigh}
\end{equation}
as anticipated by the density fluctuation contribution predicted in Eq.~(\ref{eq_fThigh}).
In contrast to Eq.~(\ref{eq_eThigh}) for the mean energy the density fluctuation term does
now correspond to the leading contribution to the numerically measured property.
This increases the range where the density fluctuation contribution can be demonstrated 
to over three decades in $x$.
Eq.~(\ref{eq_cVThigh}) is indicated by the bold lines in the main panel and the inset
of Fig.~\ref{fig_cVeps}. 
%

\paragraph*{Scaling with chain length $N$.}
We have still to clarify the scaling observed for $u=N/\geff(x) \ll 1$ in the inset of Fig.~\ref{fig_cVeps}.
%
Chains which are smaller than the thermal blob ($u \ll 1$) behave obviously as random walks 
and the density fluctuations decouple from the interaction strength $x$.
Due to this factorization the specific heat for these short chains must scale as $x^2$, 
just as for beads. This is shown in the main figure for $N=16$ (thin solid line).
Consistency with Eq.~(\ref{eq_cVThigh}) implies the scaling
$\cV(x,N) \approx x^{3/2} \rho^{1/2} h(u)$ with $h(u)$ being a universal function
scaling as $h(u) \sim u^0$ in the large-$u$ limit. Since $\cV \sim x^2$ 
and $\geff(x) \sim 1/(x\rho)$ (as shown in Sec.~\ref{sub_compress}) for $u \ll 1$,
it follows that $h(u) \approx u^{1/2}$ as confirmed by the dashed line indicated in the inset.
Hence, $\cV \sim \rho x^2 N^{1/2}$ for $u \ll 1$. 
%
%

\subsection{Chemical potential}
\label{sub_mu}

\paragraph*{Scaling of the chemical potential.}
Fig.~\ref{fig_mueps} presents the excess chemical potential per monomer, 
$y \equiv \muex/TN$,
obtained using thermodynamic integration (as explained below) for three chain lengths 
$N=1$, $64$, and $2048$ as functions of the overlap penalty $x=\overlap/T$.
As one expects, the chemical potential increases first linearly with $x$ and then levels off. 
Chain length effects are again small on the logarithmic scale chosen in the plot \cite{foot_muN}. 
For large $x$ the chemical potential becomes slightly larger for beads ($y\approx 2.64$) 
than for long chains where $y \approx 2.1$ (dash-dotted line). 
That the chemical potential of polymer chains is reduced compared to melts of unconnected 
beads is of course expected for all $x$ due to the (effectively) attractive bond potential.
For weak interactions this reduction should be described by the density fluctuation 
contribution to the free energy [Eq.~(\ref{eq_fThigh})] 
which corresponds to an excess chemical potential
\begin{equation}
y = \frac{\partial (\beta f(\beta) \rho)}{\partial \rho} 
\approx \vone(x) \rho \left( 1 - \
 \underline{ \frac{3\sqrt{3}}{\pi} \frac{(\vone(x)\rho)^{1/2}}{\be(x)^3\rho}}
+ \ \ldots \right)
\mbox{ for } x \ll 1
\label{eq_muThigh}
\end{equation}
with $v(x)$ being the second virial of unconnected beads.
The dashed line in Fig.~\ref{fig_mueps} presents the leading contribution $v(x) \rho$ for unconnected beads,
the bold line in addition the underlined connectivity contribution given in Eq.~(\ref{eq_muThigh}).
It turns out that the simple second virial approximation provides a much better fit of the data over 
the entire $x$-range than the full prediction. (The weak underestimation of the chemical potential for $x > 10$ 
must be attributed to higher order correlations relevant in this limit.) 
That the density fluctuation contribution overestimates the reduction of the chemical potential for $x > 1$ 
is in agreement with Eq.~(\ref{eq_Ginzburg}) and the Ginzburg parameters indicated in Table~\ref{tab_epsasym}. 
Hence, Eq.~(\ref{eq_muThigh}) in principle can be tested only for $x \ll 1$. Unfortunately, in this limit
the relative correction, scaling as $\sqrt{x/\rho}$, becomes too small to allow a fair test of the theory. 
A systematic variation of the density and an improved numerical accuracy of the chemical potentials measured 
are warranted to achieve this goal.

\paragraph*{Thermodynamic integration.}
We now explain how the data of Fig.~\ref{fig_mueps} have been obtained numerically.
The simple insertion method due to Widom \cite{FrenkelSmitBook} obviously becomes rapidly 
inefficient with increasing $x$, especially for longer chains. Slightly generalizing the method suggested
in \cite{MP94,WM94} we therefore have performed a thermodynamic integration \cite{FrenkelSmitBook}
\begin{equation}
\beta \muex = \int_{\lambda(\overlap)}^{1} d\lambda \frac{\la \Novghost \ra}{\lambda}
\label{eq_thermoint}
\end{equation}
over discrete values of the interaction affinity $\lambda = \exp(-\overlapghost \beta/8)$ 
characterizing the excluded volume interaction of a ghost (g) chain that is gradually inserted into 
an equilibrated system (s). 
$\la \Novghost \ra$ refers to the mean number of lattice sites where system and ghost monomer cube corners overlap
at a given interaction $\lambda$.
Generalizing the Potts spin mapping, Eq.~(\ref{eq_spinmodel}),
of the excluded volume interactions for homopolymers presented above, we use now {\em two} spin lattices,
$\sigma_\text{s}(\bm{r})$ describing (as before) the interaction of the system monomers and
$\sigma_\text{g}(\bm{r})$ the ghost chain. The spin lattices are kept at the same temperature
and are both characterized by the same (arbitrary) energy $\overlap=1$ which has to be paid 
for a complete overlap of two system monomers or two ghost monomers.
The interaction of both spins is described by
\begin{equation}
\Delta E_{sg} = \sum_{\bm{r}} \sigma_\text{s}(\bm{r}) \sum_{\bm{\delta}} J_\text{sg}(\bm{\delta}) \sigma_\text{g}(\bm{r}+\bm{\delta})
\label{eq_ghostinter}
\end{equation}
with coupling constants $J_\text{sg}(\bm{\delta}) \sim \overlapghost$ defined as in Eq.~(\ref{eq_J}) 
taken apart the energy parameter $\overlap$ which is replaced by the tuneable interaction energy $\overlapghost$.
Starting with decoupled system and ghost configurations at $\overlapghost=0$, i.e. $\lambda=1$, 
we gradually increase the interaction parameter 
up to $\overlapghost=\overlap$, i.e.~ $\lambda(\overlap) = \exp(-\overlap \beta/8)$, 
always keeping the coupled system at equilibrium. 
Monitoring the distribution of the number $\Novghost$ of overlaps between system and ghost cube corners 
we use multihistogram methods as described in \cite{MP94,WM94} to improve the precision of the integral.

The mean overlap number $\la \Novghost \ra$ (devided by $8N$) is shown in the inset of 
Fig.~\ref{fig_mueps} as a function of $\lambda$ for $N=2048$ and two inverse temperatures 
$x=3$ and $x=100$. Starting from $\lambda=1$ the overlap number decreases monotonously 
with increasing coupling between system and ghost monomers,  i.e. decreasing $\lambda$.
Interestingly, a power law behavior $\la \Novghost \ra/N \approx \lambda^{1/4}$ is found
empirically for large $x \gg 10$ (dashed line). Fitting this power law and integrating 
then analytically Eq.~(\ref{eq_thermoint}) provides a useful crosscheck of the numerical 
integration using the multihistogram analysis. This is a technically important finding, 
since the multihistogram analysis requires overlapping distributions of $\Novghost$ and 
hence much more equilibrated intermediate values $\lambda$ as indicated for $x=100$. 
A detailed explanation for the observed power law still is missing, but it is presumably
due to the systematic screening of the long range correlations of the ghost chain which
is swollen at $\lambda=1$ becoming more and more Gaussian as it feels the compression
due to the surrounding host chains \cite{DegennesBook}.

\subsection{The compressibility}
\label{sub_compress}

\paragraph*{Compressibility $g(x,N)$ and excess compressibility $\gex(x)$.}
To test the key relation Eq.~(\ref{eq_Rsgen}) announced in the Introduction we need accurate values 
for the dimensionless compressibilities $\geff(x) \equiv \lim_{N\to \infty} g(x,N)$ of asymptotically
long chains. As suggested by Eq.~(\ref{eq_gdef}), we compute first the dimensionless compressibility 
$g(x,N) = \lim_{q\to 0} G(q)$ from the low-$q$ limit of the total monomer structure factor
for different overlap penalties $x$ and chain lengths $N$  (see below for details). 
These raw data are presented in Fig.~\ref{fig_geps} as a function of $x$.
As one expects, $g(x,N)$ decreases monotonously with overlap strength $x$.
%
%
In contrast to the thermodynamic integration performed for the chemical potential [Eq.~(\ref{eq_thermoint})],
the structure factor measures the complete compressibility, not just the excess contribution.
The strong $N$-dependence visible in the plot thus is expected from the translational entropy of the chains.
As can be seen, e.g., from Eq.~(\ref{eq_fThigh}) or from the virial expansion of
polymer solutions \cite{DegennesBook}, the compressibility can be written in general as
\begin{equation}
\frac{1}{g(x,N)} =
\rho \ \frac{\partial^2(\beta f(\beta) \rho)}{\partial \rho^2} 
=
\frac{1}{N} + \frac{1}{\gex(x,N)} 
\label{eq_gNeffect}
\end{equation}
for all $x$ with $\gex(x,N)$ being the excess contribution to the compressibility
which may, at least in principle, depend on $N$ \cite{foot_muN}.
As can be seen from the inset of Fig.~\ref{fig_geps}, {\em all} compressibilities collapse,
however, on {\em one} $N$-independent master curve if one plots $1/g(x,N)-1/N$ as a function of $x$,
even the compressibilities obtained for unconnected beads ($N=1$).
Within numerical accuracy the $N$-dependence observed for $g(x,N)$ can therefore be attributed 
to the trivial osmotic contribution and the excess compressibility $\gex \sim N^0$ is thus  
identical to the compressibility $\geff(x)$ of asymptotically long chains.
The bold line indicated in the inset presents the best values of $\geff(x)$
summarized in Table~\ref{tab_epsasym}. 
These values have been obtained from the excess compressibilities for the largest chain length 
available for $x \ge 0.001$. A {\em precise} numerical determination of $\gex(x)$ becomes impossible 
for even smaller overlap penalties. We thus have used for the smallest $x$-values sampled
the theoretical prediction
\begin{equation}
\frac{1}{\geff(x)} 
\approx \vone(x) \rho 
\left(1 - \
\underline{\frac{3\sqrt{3}}{2\pi} \frac{(\vone(x)\rho)^{1/2}}{\be(x)^3\rho}}
\ldots \right)
\mbox{ for } x \ll 1
\label{eq_gThigh}
\end{equation}
due to the free energy [Eq.~(\ref{eq_fThigh})] postulated in the Introduction.
The prefactor $v(x)\rho$ representing the bare monomer interaction is indicated
by the dashed line in the main panel of Fig.~\ref{fig_geps}.
Hence, $\geff(x) \approx 1/(8x\rho)$ for weak interactions, 
i.e. the compressibility increases linearly with temperature.
The underlined term is the leading correction due to the density fluctuation contribution 
to the free energy. It implies that the excess compressibilities for polymer melts and 
unconnected beads cannot be completely identical. However, as before for the chemical potential, 
the difference is far too small to be measurable in the limit where Eq.~(\ref{eq_gThigh}) applies.
Although this result is unfortunate from the theoretical point of view,
the data collapse observed in the inset suggests that it is acceptable to numerically estimate
the long chain compressibility $\geff(x)$ by computing the structure factors of rather short chains.

%
%

\paragraph*{Total monomer structure factor.}
We now turn to the total structure factor $G(q)$ shown in Fig.~\ref{fig_Sq} to explain
how the compressibilites $g(x,N)$ have been obtained. Only chains of length $N=2048$
are presented for clarity. The total monomer structure factor is obtained by computing
$G(q) = \frac{1}{\nmon} \la [\sum_n \cos(\bm{q}\cdot\bm{r}_n)]^2 + [\sum_n \sin(\bm{q}\cdot\bm{r}_n)]^2 \ra$ 
where the sums run over all the $\nmon$ monomers of the box and the wavevectors are commensurate 
with the cubic box of linear length $L$. Since the smallest possible wavevector is $2\pi/L$,
it thus is important to have large box sizes to scan over a sufficiently important $q$-range
allowing a reasonable determination of $g(x,N)$.
Note that around and above $q \approx 2$ monomer structure and lattice effects become important.
Since only smaller wavevectors are of interest if one is interested 
in universal physical behavior, we will focus below on wavevectors $q < 1$.
For comparison, we have also included the single chain form factor 
$F(q) = \frac{1}{N} \la [\sum_{n=1}^N \cos(\bm{q}\cdot\bm{r}_n)]^2 + 
[\sum_{n=1}^N \sin(\bm{q}\cdot\bm{r}_n)]^2 \ra$
\cite{DoiEdwardsBook} for $x=0.001$ (bold line).
Note that the qualitative shape of $F(q)$ 
--- decaying monotonously with $q$ from its maximum value $F(q=0)=N$ ---
depends very little on the temperature (not shown).
We remind that the ``random phase approximation" (RPA) formula 
\cite{DegennesBook,DoiEdwardsBook}
\begin{equation}
\frac{1}{G(q)} = \frac{1}{F(q)} + \frac{1}{\geff(x)} 
\label{eq_RPA}
\end{equation}
relates the total structure factor to the measured form factor.
Eq.~(\ref{eq_RPA}) is of course consistent with Eq.~(\ref{eq_gNeffect}) in the $q\to 0$ limit.
It allows to directly fit for the excess compressibility $\gex(x)$ 
using the measured structure factor $G(q)$ and form factor $F(q)$,
at least in the $x$-range where the RPA approximation applies.
As may be seen from the figure, $G(q)$ indeed decreases systematically with $x$, 
i.e. with decreasing $\geff(x)$. For large temperatures ($x \le 3$) it also decays monotonously
with $q$, again in agreement with Eq.~(\ref{eq_RPA}). 
Interestingly, this becomes qualitatively different for larger excluded volume interactions 
($x > 3$) where the total structure factor is essentially constant (in double-logarithmic coordinates), 
very weakly {\em increasing} monotonously with $q$. 
The RPA formula apparently does not apply in this limit in agreement with Eq.~(\ref{eq_Ginzburg}).
Fortunately, this is of no concern for our main purpose --- to compute $\geff(x)$ ---
since in precisely this limit the compressibility is readily obtained from a broad plateau
(even for much smaller boxes) which in addition becomes chain length independent,
as we have already seen from the inset of Fiq.~(\ref{fig_geps}). 
Using boxes with $L=256$ it is possible to directly measure the plateau values for $x \le 0.3$.
For smaller $x$ we have simulated boxes with $L=512$ containing $\nmon \approx 8.4 \cdot 10^6$ monomers
and corresponding to a smallest wavevector $q \approx 0.01$. 
This box size becomes again insufficient for the largest temperatures we have simulated,
as shown in Fig.~\ref{fig_Sq} for $x=0.001$ (dashed line). It is for these values where the 
RPA formula, Eq.~(\ref{eq_RPA}), allowing to fit the {\em deviation}
from the (barely visible) plateau, has been particulary useful.

\paragraph*{Approximated RPA formula.}
We finally note that in the intermediate wavevector regime 
(where $q$ corresponds to distances much smaller than the radius of gyration
and much larger than the monomer size) the general RPA Eq.~(\ref{eq_RPA}) may be written as
\begin{equation}
\frac{1}{G(q)} = \frac{1}{\geff(x)} + \frac{1}{12} \be^2(x) q^2 = \frac{1}{\geff(x)} \left(1 + (q \xi)^2 \right) 
\label{eq_RPAapprox}
\end{equation}
which justifies the definition given in Eq.~(\ref{eq_xidef}) for the correlation length 
of the density fluctuations $\xi$. 
Here we have used that the form factor becomes $F(q) \approx 12/\be^2 q^2$ \cite{DoiEdwardsBook}. 
This assumes that corrections to Gaussian chain statistics may be ignored \cite{WBJSOMB07,BJSOBW07} 
and that finite chain size effects are negligible.
From the numerical point of view the approximated RPA Eq.~(\ref{eq_RPAapprox})
has the disadvantage that the effective bond length $\be(x)$ needs to be determined first. 
As shown in Fig.~\ref{fig_Sqxi}, it has the advantage that it allows for an additional 
test of the values of $\geff(x)$ and $\be(x)$ indicated in Table~\ref{tab_epsasym}.
The main panel presents the rescaled structure factor $G(q)/\geff(x)$ for chains of length $N=8192$
as a function of $Q \equiv q\xi$ with $\xi$ being obtained from $\geff(x)$ using Eq.~(\ref{eq_xidef}).
All data collapse on the master curve $1/(1+Q^2)$ indicated by the bold line provided that the
wavevector $q$ remains sufficiently small and no local physics is probed.
That the used compressibilities are accurate is emphasized further in the inset
where $\geff(x)/G(q)-1$ is plotted as a function of $Q^2$ using only sufficiently small wavevectors $q$. 
According to Eq.~(\ref{eq_RPAapprox})
all data should collapse on the bisection line in double-logarithmic coordinates 
if the correct compressibilities are used. This is indeed the case.
Unfortunately, even this rather precise method still does not allow to demonstrate
the density fluctuation contribution in Eq.~(\ref{eq_gThigh}) for $x \ll 1$ 
since the same scaling collapse is obtained for the simple choice $1/\geff(x) \equiv v(x) \rho$.
Please note the weak deviations visible for $x=1$ which is due to the breakdown of the RPA formula
for large $x$ mentioned above.

\subsection{Bond properties}
\label{sub_bonds}

Up to now we focused on some thermodynamic features of the soft BFM model,
i.e. on large-scale properties. Turning to configurational issues we begin
by characterizing local-scale features of the algorithm. (Readers only interested
in universal properties may wish to skip this paragraph.) 

\paragraph*{Mean bond length.}
By definition of our version of the BFM algorithm the bond length is allowed to fluctuate 
strongly between $2$ and $\sqrt{10}$. One expects that switching on the excluded volume interaction, 
i.e. decreasing the temperature, will suppress large bonds due to the increasing pressure.
The mean bond length commonly is characterized by the root-mean-square length $l=\la \bm{l}^2\ra^{1/2}$.
(Other moments would yield similar results.) The mean bond length rapidly becomes ($N > 20$)
chain length independent \cite{Paul91a}. As a function of overlap penalty $x$ the bond length shows 
a monotonous decay between $x \approx 3$ and $x \approx 20$ as can be seen from Fig.~\ref{fig_be}.
As other local properties, $l$ becomes constant in the small-$x$ and large-$x$ limits (dashed horizontal lines).
The value $l(x=0)=2.718$ gives the lower bound for the effective bond length $\be(x)$ of asymptotically
long chains (stars) obtained below.

\paragraph*{Mean bond angle and local chain rigidity.}
Defining the bond angle $\theta$ between two subsequent bonds by the scalar product 
$\cos(\theta) = \bm{e}_n \cdot \bm{e}_{n+1}$ of the normalized bond vectors
$\bm{e}_i = \bm{l}_i/|\bm{l}_i|$, the {\em local} chain rigidity may be characterized by
$\la \theta \ra$ and $\la \cos(\theta) \ra$. 
Note that $\la \theta \ra$ and  $\la \cos(\theta) \ra$ can be regarded as chain length independent,
just as the mean bond length. As can be seen from Table~\ref{tab_epsasym}, 
the local rigidity is negligible for $x \ll 1$, i.e. $\la \theta \ra \approx 90^{\circ}$
and $\la \cos(\theta)) \ra \approx 0$ due to the symmetry of the distribution $p(\theta)$ 
with respect to $90^{\circ}$.
The rigidity then increases around $x \approx 1$ and becomes constant again for large $x$ 
where $\la \theta \ra \approx 82.2^{\circ}$ and $\la \cos(\theta) \ra \approx 0.106$. 
The increase of the local rigidity for larger excluded volume interactions is of course expected 
due to the suppression of immediate backfoldings corresponding to bond angles $\theta > 143^{\circ}$ \cite{WPB92}. 
The distribution $p(\theta)$ therefore becomes lopsided towards smaller $\theta$ (not shown).
It is well known \cite{DoiEdwardsBook} that for chains characterized by the 
``freely rotating" (FR) chain model such a local rigidity would lead to an effective bond length
$\be(x) = l(x) \sqrt{c_\text{FR}}$ with $c_\text{FR} = (1+\la \cos(\theta) \ra)/(1-\la \cos(\theta) \ra)$.
This simple model, indicated by the crosses in Fig.~\ref{fig_be}, yields a qualitatively reasonable
trend (monotonous increase of the effective bond length at $x\approx 1$) but fails 
to fit the directly measured effective bond lengths quantitatively.

\subsection{Chain and segment size}
\label{sub_chainsize}

\paragraph*{Total chain size $\Rend(N)$.}
\label{para_RN}
One way to characterize the total chain size is to measure the second moment of the 
chain end-to-end distance $\Rend^2(x,N) = \la (\bm{r}_N-\bm{r}_1)^2 \ra$. 
(Other moments yield similar behavior \cite{WBM07}.)
We consider the effective bond length $b(x,N) \equiv \Rend(x,N)/\sqrt{N-1}$ to compare 
the measured chain size with the ideal chain behavior which is commonly taken as granted
\cite{DegennesBook,DoiEdwardsBook,Flory49} and which is the basis of our
perturbation calculation. 
We use the notation $\be(x)\equiv \lim_{N\to\infty} b(x,N)$ 
for the effective bond length of asymptotically long chains \cite{foot_asymlimit}. 
The effective bond lengths $b(x,N)$ for $N=64$ and $N=2048$ and 
the asymptotic limit $\be(x)$ --- obtained by extrapolation as described below --- 
are presented in Fig.~\ref{fig_be} as functions of $x$. 
Obviously, $b(x,N) \to l(x=0)$ for all $N$ in the small-$x$ limit. 
$b(x,N)$ increases then in the intermediate $x$-window before it levels off at $x\approx 10$.
The swelling due to the excluded volume interaction is the stronger the larger the chain length,
i.e. $b(x,N)$ increases monotonously with $N$. This swelling therefore cannot be attributed
to a {\em local} persistence length as described, e.g., by the freely-rotating chain model.

The chain length effects can be seen better in Fig.~\ref{fig_ReN} where we have plotted
$b(x,N)$ for several penalties $x$ as a function of $t=1/\sqrt{N-1}$.
The choice of the horizontal axis is motivated by Eq.~(\ref{eq_Rsmelt}) suggesting 
the linear relation
\begin{equation}
b^2(x,N) \approx \be^2(x) \ (1 - c(x) \ce(x) t) 
\label{eq_RNfit}
\end{equation}
for $u=N/g \gg 1$ 
with $\ce(x) \equiv \sqrt{24/\pi^3}/\rho \be(x)^3$ being the swelling coefficient defined in Sec.~\ref{sec_intr}.
$c(x)$ is an additional numerical prefactor of order unity which has been introduced in agreement with Eq.~(19) 
of Ref.~\cite{WBM07}. The reason for this coefficient is that the corrections to Gaussian behavior differ
slightly for internal chain segments [as described by Eq.~(\ref{eq_Rsmelt})] and the total chain size
which is characterized in Fig.~\ref{fig_ReN}. 
It has been shown that $c \to 1.59$ for large $N$ \cite{WBM07}.
However, since this value corresponds to the limit of a very slowly converging integral \cite{WBM07} 
it is better to use Eq.~(\ref{eq_RNfit}) as a two-parameter fit for $\be(x)$ and $c(x)$
and to crosscheck then whether the fitted $c$ is of order unity.  
As shown in the figure for three overlap penalties, this method can be used reasonably for overlap penalties 
as low as $x \approx 0.1$, albeit with decreasing $x$ systematically underestimating the ``true" $\be$-values 
indicated in Table~\ref{tab_epsasym}. Please note that $N/g \approx 400$ for $x=0.1$ and $N=8192$. 
Chains with $N \gg 8192$ would be required to use this method for even smaller $x$.
In this limit it is better to use as a simple first step the value $b(x,N=8192)$ of the largest chain 
length simulated as a (rather reasonable) lower bound for $\be(x)$.

\paragraph*{Segment size $R(s)$.}
\label{para_Rs}
As we have already stressed in Ref.~\cite{WBM07}, it is technically better to extrapolate
for the effective bond length $\be$ using the distribution $R(s)$ of the mean-squared
size of segments of curvilinear arc-length $s$ defined in Sec.~\ref{sec_intr}.
It follows from $c > 1$ that the total chain ratio $b^2(x,N)$ converges less rapidly to the 
asymptotic Gaussian behavior as $R(s)^2/s$. (See Fig.~4 of Ref.~\cite{WBM07}.)
The ratio of $R^2(s)/s$ as a function of $s$ is plotted in the inset of Fig.~\ref{fig_Rsscal}
for $N=2048$ and for several $x$. As can be seen, it increases systematically with 
segment length $s$. The swelling levels off for large $s$, but rather gradually.
Therefore it would not be appropriate to identify the maximum around $s \approx N$ 
as the asymptotic plateau. This again would yield an {\em underestimation} of $\be(x)$. 
A more precise method to obtain $\be(x)$ uses the predicted correction, Eq.~(\ref{eq_Rsmelt}),
to the Gaussian limit. We recommend to plot, using double-logarithmic coordinates,
$1-R(s)^2/\be^2s$ as a function of $\ce/\sqrt{s}$ and to tune $\be(x)$ until the data for 
intermediate chain segments with $g \ll s \ll N$ collapses on the bisection line.
This {\em one}-parameter fit yields good estimates down to $x=0.01$ where $N/g \approx 40$. 
Since the corresponding plot is very similar to Fig.~5 of Ref.~\cite{WBM07},
it is not reproduced here.
We rather show in Fig.~\ref{fig_Rsscal} a scaling plot motivated by the key relation, 
Eq.~(\ref{eq_Rsgen}), which uses our best values of $g(x)$ and $\be(x)$
for asymptotically long chains (Table~\ref{tab_epsasym}).
The data collapse on the theoretical prediction (bold line) is remarkable,
especially considering that $u$ covers seven orders in magnitude. 
The ``Fixman limit", Eq.~(\ref{eq_Rsfixm}), for $u \ll 1$ (dashed line)
fits the data with the smallest overlap penalty $x \approx 0.001$
confirming the chosen value of $\be$. 
The limiting behavior for $u \gg 1$ [Eq.~(\ref{eq_Rsmelt})], characterizing an incompressible 
melt of thermal blobs of length $g$, is indicated by the dash-dotted line.

\paragraph*{Predicting the effective bond length.}
\label{para_be}
Up to now, we have used theory to improve the {\em fit} of $\be(x)$, rather than to 
{\em predict} it from the thermodynamic properties and local model features such as 
the bond length $l(x)$. The increase of the effective bond length for weakly interacting 
and asymptotically long polymer melts has been calculated long ago by Edwards 
[see Eq.~(5.55) of Ref.~\cite{DoiEdwardsBook} or Eq.~(11) of Ref.~\cite{WBM07}]. 
Reformulated using our notations 
and substituting the bare excluded volume parameter $v(x)$ by 
$1/g\rho$ \cite{ANS05a,WBM07,foot_veff}
his result reads
\begin{equation}
\be^2 = l^2 \left( 1 + \frac{\sqrt{12}}{\pi} \Gi \right)
\mbox{ with } \Gi = \frac{1}{\sqrt{g} \br^3\rho}
\label{eq_bepredict}
\end{equation}
where $\br$ is the bond length of the unperturbed reference chain of the calculation
and $\Gi(\br,g)$ the relevant Ginzburg parameter quantifying the strength of the interaction
acting on a chain segment of length $s=g$ (see Eq.~(6) of Ref.~\cite{WBM07}).
Since $\Gi$ becomes small for large compressibilities $g(x)$, one expects good agreement 
with our data for small $x$.
The question is now what actually might be the best reference bond length
to allow a prediction over the broadest possible $x$-range.
The simplest choice to associate $\br$ with the bond length $l(x)$
yields the dash-dotted line indicated in Fig.~\ref{fig_be}.
As can be seen, this choice of $\br$ allows a reasonable prediction
only up to $x \approx 0.01$. 
The predictive power of Eq.~(\ref{eq_bepredict}) can be considerably improved over 
nearly two decades up to $x \approx 1$ if one applies the formula iteratively starting with
$\br =l$ and using the effective bond length obtained as input for the Ginzburg parameter
($\be \to \br$) in the next step. This recursion relation converges rapidly as shown by the
bold line indicated in Fig.~\ref{fig_be} obtained after 20 iterations. 
This iterative renormalization of the bond length of the reference chain and the associated 
Ginzburg parameter has been suggested by Muthukumar and Edwards \cite{ME82}. 
Essentially the same result is obtained up to $x \approx 1$ 
if one sets directly $\br=\be$ using the {\em measured} effective bond length (not shown), 
i.e. these values correspond to the fix-point solution of Eq.~(\ref{eq_bepredict}).
The Ginzburg parameters $\Gi$ obtained using the measured $\be(x)$ are listed in Table~\ref{tab_epsasym}.
Note that $\Gi < 0.34$ for $x < 1$ where Eq.~(\ref{eq_bepredict}) fits our data nicely.
The fix-point solution of Eq.~(\ref{eq_bepredict}) does not capture correctly the leveling off 
of $\be(x)$ setting in above $x \approx 1$. Since the Ginzburg parameter becomes there of order one
(Table~\ref{tab_epsasym}), this is to be expected.
In summary, we have shown that the iteration of Eq.~(\ref{eq_bepredict}) allows a good 
prediction for $\be(x)$ for $x < 1$ such that $\Gi \ll 1$. If reliable values for the
compressibility $g(x)$ are available [by means of the extrapolation method implied by
Eq.~(\ref{eq_gNeffect})], this is the method of choice if one cannot afford to 
simulate very long chains.
%

\subsection{Bond-bond correlation function}
\label{sub_Ps}

\paragraph*{Motivation and theoretical prediction.}
We return now to the deviations from Flory's ideality hypothesis 
predicted in Eq.~(\ref{eq_Rsgen}) for the mean-square segment size $R^2(s)$. 
As we have seen above (Fig.~\ref{fig_Rsscal}), this property requires to substract 
a large Gaussian contribution $\be^2s$ from the measured $R^2(s)$ to 
demonstrate the existence and the scaling of the deviations.
Unfortunately, this requires as a first step the precise determination of the 
effective bond length $\be(x)$ for asymptotically long chains which might not always be available.
Indeed we have used in the preceeding Sec.~\ref{sub_chainsize} the fact that the scaling 
of Eq.~(\ref{eq_Rsgen}) critically depends on this accurate value to {\em improve} the 
estimation of the effective bond length $\be(x)$ for asymptotically long chains. 
Hence, it would be nice to demonstrate directly the scaling implied by our key 
prediction {\em without} any tuneable parameter. 
The trick to achieve this is similar to our demonstration of the density fluctuation 
contributions to the free energy, Eq.~(\ref{eq_fThigh}), presented in Sec.~\ref{sub_energyfluctu}:
We consider the {\em curvature} of $R^2(s)$, i.e. its second derivative with respect to $s$,
to eliminate the large Gaussian contribution. In principle this can be achieved by fitting
$R^2(s)$ by a sufficiently high polynomial whose second derivative with respect to $s$ then is
compared to the theory. Following \cite{WMBJOMMS04,WBM07} we use a more direct numerical 
route where we compute the well-known bond-bond correlation function 
$P(s)  \equiv \la \bm{l}_{m=n+s} \cdot \bm{l}_n \ra / l^2$
with $\bm{l}_i =\bm{r}_{i+1}-\bm{r}_i$ denoting the bond vector between two adjacent monomers
$i$ and $i+1$ and $l^2$ the mean-square bond length (Sec.~\ref{sub_bonds}).
The average is performed as before over all chains and all pairs of monomers $(n,m+s)$
possible in a chain of length $N$.
We use this definition rather than the more common first Legendre polynomial 
$\la \bm{e}_n \cdot \bm{e}_m \ra$ 
since it allows to relate the bond-bond correlation function to the segment size by
\begin{equation}
P(s) = \frac{1}{2l^2} \frac{d^2}{ds^2} R^2(s).
\label{eq_PsRsrelated}
\end{equation}
This formula is obtained from 
$ \la \bm{l}_n \cdot \bm{l}_m \ra \approx \la \partial_n \bm{r}_n \cdot \partial_m \bm{r}_m \ra
= - \partial_n \partial_m \la (\bm{r}_n - \bm{r}_m)^2\ra/2$.
Using Eq.~(\ref{eq_PsRsrelated}) the key prediction, Eq.~(\ref{eq_Rsgen}), 
implies for the bond-bond correlation function
\begin{equation}
P(s) = \frac{\cP}{g(x)^{3/2}}
\left( \frac{4}{\sqrt{u}} - 4 \sqrt{2\pi} e^{2u} \mbox{erfc}(\sqrt{2u}) \right)
\label{eq_Psgen}
\end{equation}
where we have introduced the coefficient $\cP = \ce (\be/l)^2/8$.
Eq.~(\ref{eq_Psgen}) corresponds to the limiting behavior
\begin{equation}
P(s)   \approx \frac{\cP}{g^{3/2}} \frac{4}{\sqrt{u}} 
\label{eq_Psfixm} 
\end{equation}
for small reduced arc-lengths $u \ll 1$. The explicit compressibility dependence drops out
in the opposite limit ($u \gg 1$) where the bond-bond correlation function becomes
\begin{equation}
P(s) \approx \cP/s^{3/2}, 
\label{eq_Psmelt}
\end{equation}
in agreement with Eq.~(\ref{eq_Rsmelt}). 
Please note that $\cP$ depends implicitly on the compressibility.
(Obviously, both asymptotic behaviors could have been obtained directly from the corresponding 
limits for $R^2(s)$, Eqs.~(\ref{eq_Rsfixm}) and (\ref{eq_Rsmelt}).)

\paragraph*{Numerical confirmation.}
The bond-bond correlation function $P(s)$ for different overlap penalties $x$ is presented
in Fig.~\ref{fig_Ps} for chains of length $N=2048$. As can be seen from the unscaled data
shown in the inset, $P(s)$ approaches a power law with exponent $\omega=1/2$ (dashed line)
in the limit of weak overlap penalties in agreement with Eq.~(\ref{eq_Psfixm}). 
For $x \ge 1$ our data is compatible with an exponent $\omega=3/2$ (dash-dotted line)
as suggested by Eq.~(\ref{eq_Psmelt}). Hence, we have demonstrated without any tunable parameter 
that Flory's ideality hypothesis is systematically violated for all segment lengths $s$ and 
all overlap penalties $x$.

We consider now the prefactors and the scaling with $x$.
As suggested by Eq.~(\ref{eq_Psgen}), the main figure presents
$P(s)/(\cP/g(x)^{3/2})$ as a function of the reduced arc-length $u=s/g(x)$
using the dimensionless compressibilities $g(x)$ and effective bond lengths $\be(x)$ from
Table~\ref{tab_epsasym}. The data collapse is remarkable as long as $1 \ll s \ll N$. 
The relation Eq.~(\ref{eq_Psgen}) is indicated by the bold line; 
it is in perfect agreement with the simulation data \cite{foot_Ps}. 
The asymptotic power law behavior 
with exponents $\omega=1/2$ for $u \ll 1$ and $\omega=3/2$ for $u \gg 1$ is shown by the dashed and 
dash-dotted lines, respectively. As predicted by Eq.~(\ref{eq_Psmelt}), one recovers the power law 
$P(s) = \cP/s^{3/2}$ --- already observed for incompressible melts \cite{WMBJOMMS04,WBM07} --- 
for scales larger than the thermal blob irrespective of the blob size $g$. 
This demonstrates that the exponent $\omega=3/2$ is not due to {\em local} physics on the monomer scale,
since for $s \gg g \gg 1$ distances much larger than the monomer or even the thermal blob are probed.
%

\section{Conclusion}
\label{sec_conc}

\begin{figure}
\centering
\includegraphics[width=12cm]{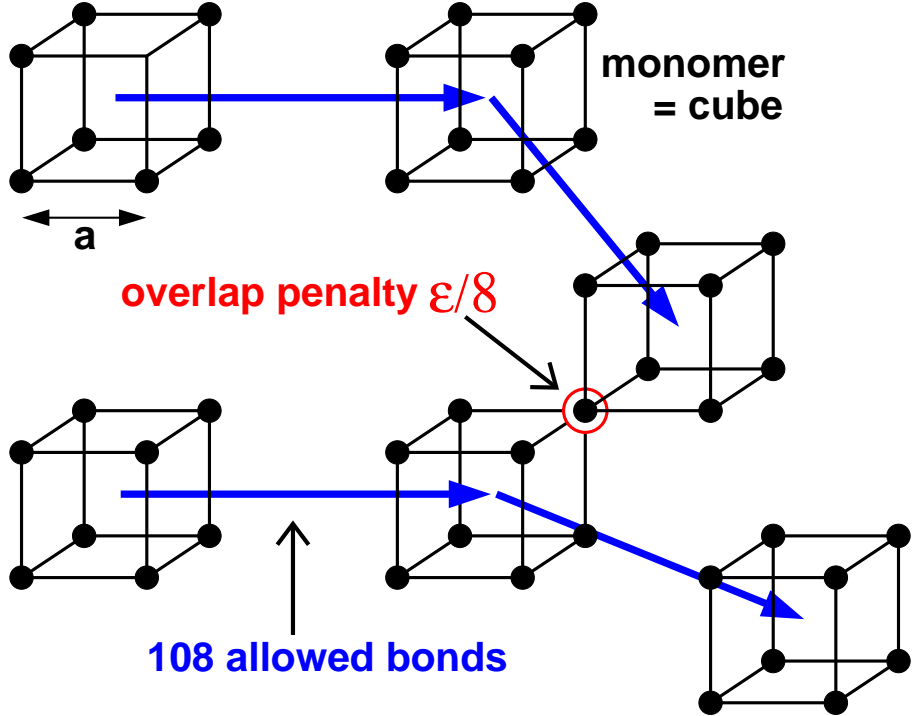}
\caption{Sketch of the bond-fluctuation model (BFM) with finite excluded volume penalty. 
The BFM algorithm represents monomers by cubes of length $a$ on a simple cubic lattice 
(of lattice constant $a$) which are connected by a set of allowed bond vectors given 
by Eq.~(\ref{eq_bondset}). Two short chains of length $N=3$ are shown.
The classical BFM model \cite{BFM,Deutsch,BWM04} assumes that all lattice sites are at most occupied once.
We relax this constraint and penalize double occupation by a {\em finite} interaction energy $\overlap$ 
which has to be paid if two cubes totally overlap. 
A corresponding fraction of the energy penality is associated with a partial monomer overlap, 
as sketched in the figure for two cube corners occupying the same lattice site.
Varying systematically the ratio $x=\overlap/T$ allows us to put to a test various theoretical results 
obtained by perturbation calculation \cite{WMBJOMMS04,BJSOBW07,WBM07,papPs} for flexible polymer chains 
in the melt.
\label{fig_BFMsketch}
}
\end{figure}

\begin{figure}
\centering
\includegraphics[width=14cm]{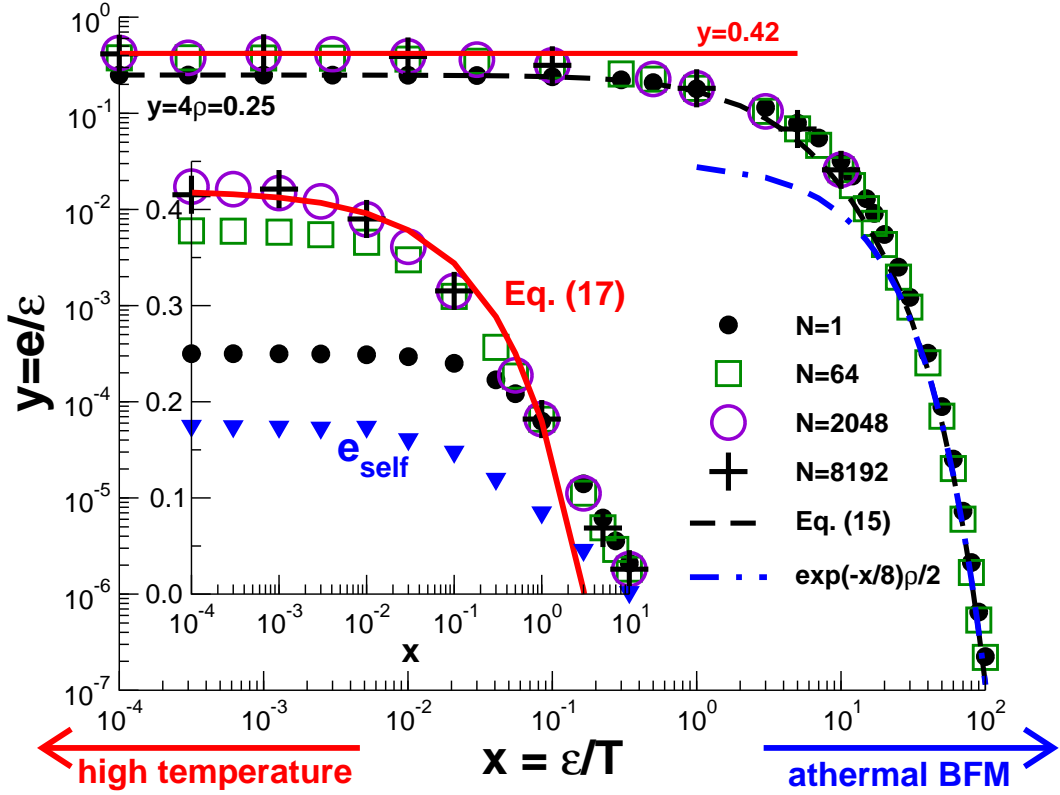}
\caption{Reduced mean overlap energy per monomer $y=e/\overlap$ as a function of 
the overlap penalty $x = \overlap/T$ for several chain lengths $N$ as indicated. 
The energy decreases monotonously with increasing $x$. The decay becomes Arrhenius-like for $x \gg 10$ (dash-dotted line). 
The dashed line indicates the energy predicted from the second virial of soft BFM beads, Eq.~(\ref{eq_evirial}). 
The main figure demonstrates the weak chain length dependence on logarithmic scales, 
especially for strong excluded volume interactions ($x > 1$). 
Inset: Same data plotted with linear vertical axis emphasizing the higher mean energy 
for long polymers ($N > 64$) for $x \ll 1$ caused by a self-energy contribution
$\eself/\overlap \approx 0.18$. The self-energies are indicated by the triangles.
The bold line shows the temperature dependence predicted by Eq.~(\ref{eq_eThigh}).
\label{fig_Eeps}
}
\end{figure}

\begin{figure}
\centering
\includegraphics[width=14cm]{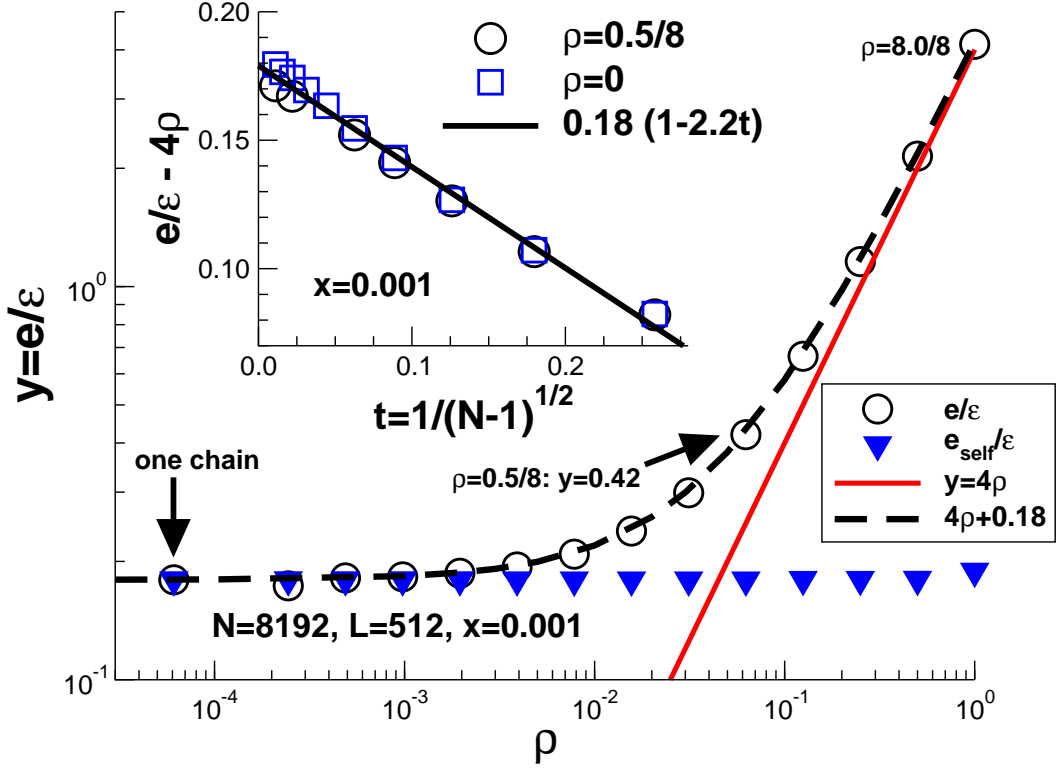}
\caption{Reduced mean energy $e/\overlap$ (spheres)
and self-energy $\eself/\overlap$ (triangles) as functions of 
the number density $\rho$ for $N=8192$, $L=512$ and $x = 0.001$.
As shown by the dashed line,
$e(\rho)$ is a superposition of the mean field energy $4\rho$ and 
the (essentially) constant self-energy $\eself/\overlap \approx 0.18 N^0 x^0 \rho^0$.
Inset: $e/\overlap - 4\rho$ as a function of chain length $1/\sqrt{N-1}$
for our reference density $\rho=0.5/8$ and for a single chain ($\rho=0$).
The linear slope (bold line) is expected from the return probability of
Gaussian chains.
\label{fig_Erho}
}
\end{figure}

\begin{figure}
\centering
\includegraphics[width=14cm]{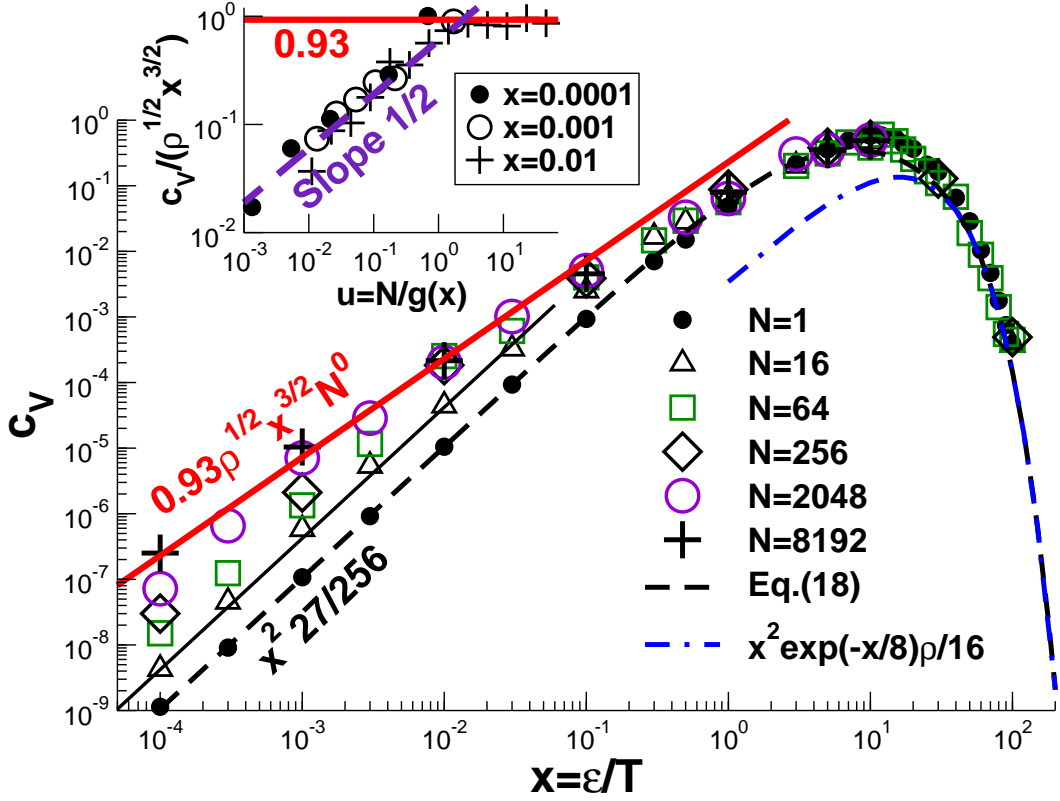}
\caption{Specific heat per bead $\cV$ {\em vs.} $x$ for chain length $N$ as indicated. 
The dashed line indicates the energy fluctuations predicted from the second virial, 
Eq.~(\ref{eq_cVvirial}), which fits nicely the data of soft BFM beads ($N=1)$ over six decades. 
While the chain length appears not to matter for strong excluded volume interactions, 
the energy fluctuations are found to increase strongly with $N$ for $x \ll 1$.
For short chains we observe $\cV \sim \rho N^{1/2} x^2$ as can be seen for $N=16$
(thin solid line).
The chain length effect saturates for long chains 
where $\cV \approx \rho^{1/2} x^{3/2} N^0$ (bold line)
in agreement with Eq.~(\ref{eq_cVThigh}).
Inset: $\cV /(\rho^{1/2} x^{3/2})$ as a function of the reduced chain length $u=N/g(x)$ with
$g(x)$ being the dimensionless compressibility (Table~\ref{tab_epsasym}).
\label{fig_cVeps}
}
\end{figure}

\begin{figure}
\centering
\includegraphics[width=14cm]{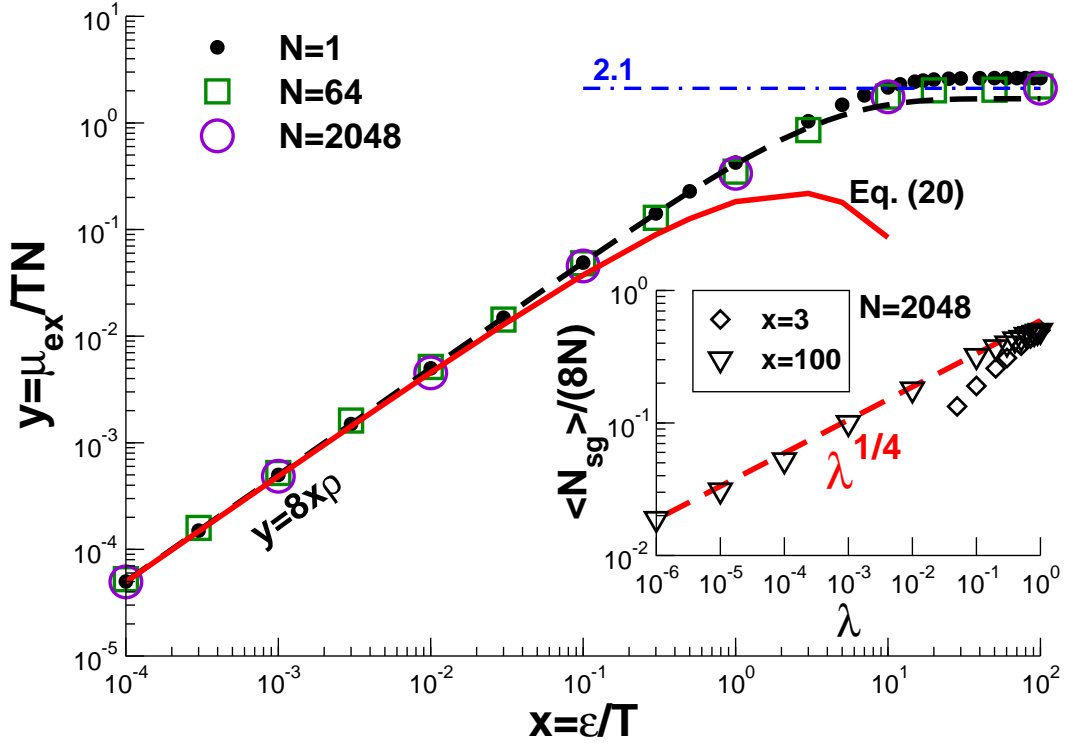}
\caption{Excess chemical potential $y=\muex/TN$ as a function of the
inverse temperature $x=\overlap/T$. Increasing linearly (dashed line) for small 
$x$ it levels off for large $x \gg 1$ (dash-dotted line). 
The dashed line indicates the simple second virial approximation $y \approx v(x) \rho$
for unconnected beads, fitting successfully the data below $x \approx 1$. The bold line 
corresponds to the high temperature prediction Eq.~(\ref{eq_muThigh}) taking into 
account the density fluctuation contribution induced by the chain connectivity.
Inset:
The chemical potential has been obtained by thermodynamic integration over the 
excluded volume interaction of an inserted ghost chain generalizing the method
suggested in Ref.~\cite{MP94}. 
The mean number of lattice sites where monomers and ghost monomers overlap, 
$\la \Novghost \ra$,
is presented for $N=2048$ as a function of 
$\lambda = \exp(-\overlapghost/8T)$ for $x=3$ and $x=100$.
A power law increase of $\la \Novghost \ra$ is found for large $x$ (dashed line).
\label{fig_mueps}
}
\end{figure}

\begin{figure}
\centering
\includegraphics[width=14cm]{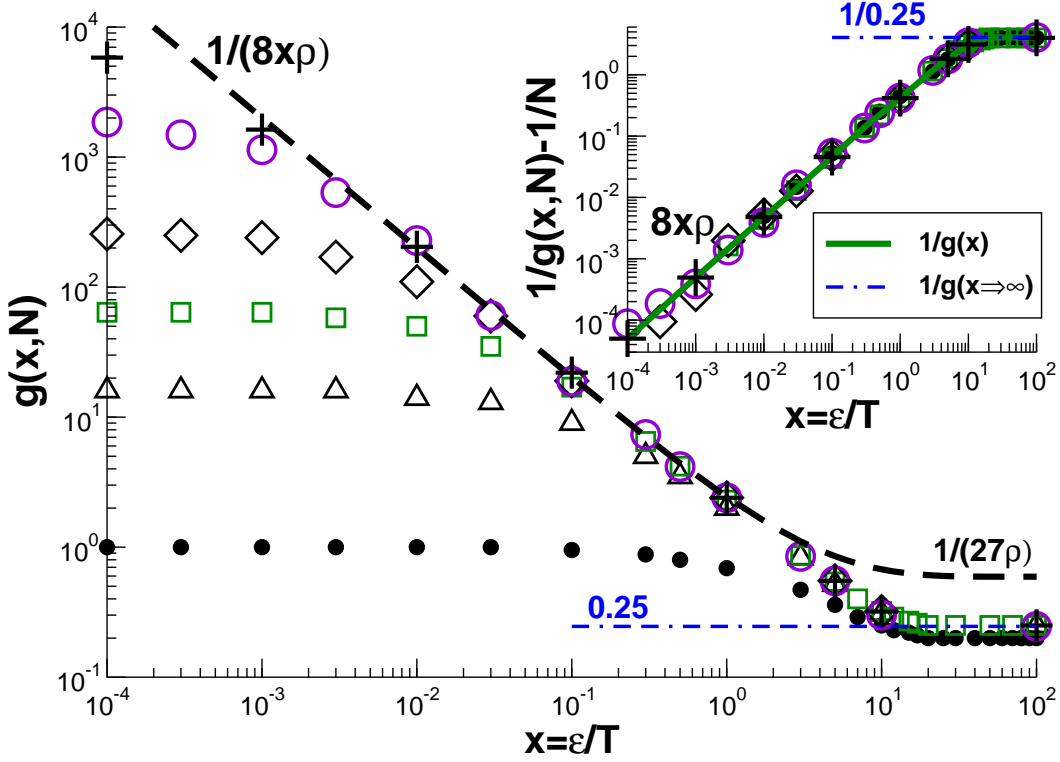}
\caption{Dimensionless compressibility $g(x,N)$ as a function of $x$ for 
different chain lengths $N$ using the same symbols as in Fig.~\ref{fig_cVeps}.
Main panel: Raw data as obtained from the low-wavevector limit of the structure factor.
Chain length effects become irrelevant for $x \ge 0.1$ if $N \ge 64$ 
and for $x > 0.001$ if $N \ge 2048$.
The data are compared to the simple second virial approximation $1/v(x) \rho$ (dashed line) 
which reduces to $1/(8x\rho)$ for $x \ll 1$.
As one expects, the compressibility levels off for large $x$ and becomes identical to the value 
$g \approx 0.25$, known for the classical BFM \cite{WBM07} (dash-dotted line).
%
%
Inset: As suggested by Eq.~(\ref{eq_gNeffect}) the excess part of the inverse compressibility
$1/g(x,N)-1/N$ becomes chain length independent, i.e. the data points for all $N$ collapse.
The master curve indicated by the bold line corresponds to the long chain limit
$g(x) = \lim_{N\to\infty} g(x,N)$ indicated in Table~\ref{tab_epsasym}. 
%
%
%
\label{fig_geps}
}
\end{figure}

\begin{figure}
\centering
\includegraphics[width=14cm]{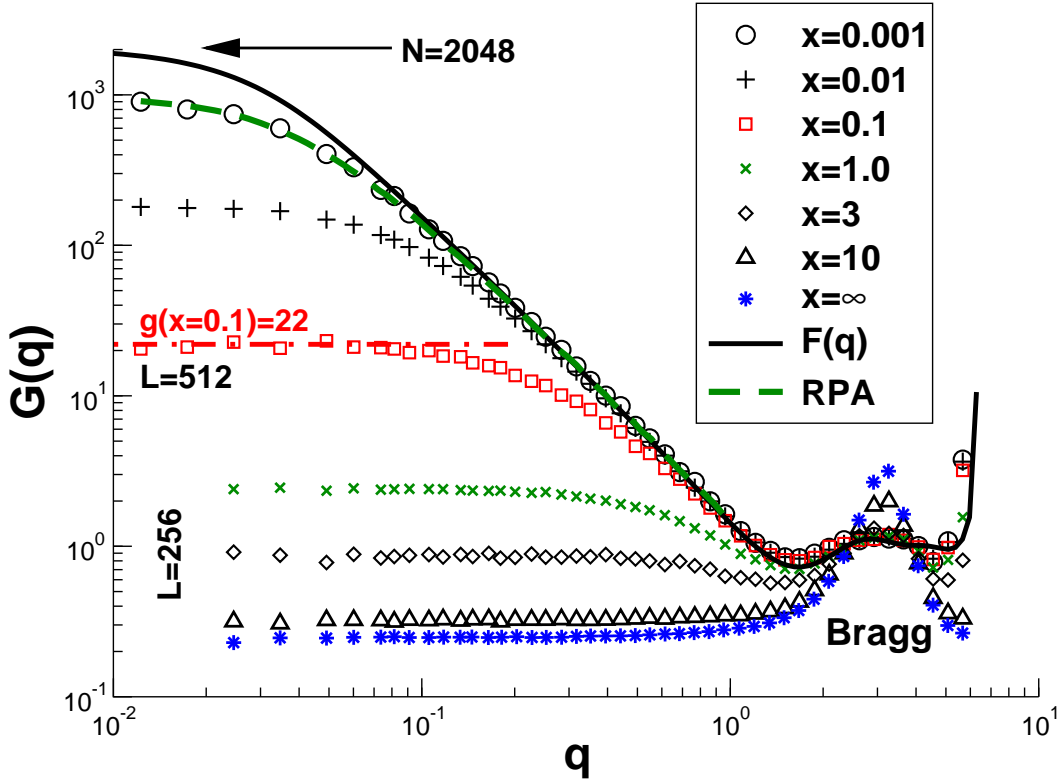}
\caption{Total structure factor $G(q)$ as a function of wavevector $q$ for $N=2048$
for different overlap penalties $x=\overlap/T$ as indicated. 
For comparison, we have also included the single chain form factor $F(q)$ for $x=0.001$. 
The low-wavevector limit of the structure factor is used to determine the dimensionless 
compressibility $g(x,N)$  [Eq.~(\ref{eq_gdef})]. 
%
%
Only for $x \le 3$ does the structure factor decay monotoneously with $q$
as suggested by the RPA formula, Eq.~(\ref{eq_RPA}).
$G(q)$ becomes essentially constant for smaller temperatures
except for wavevectors corresponding to the first sharp diffraction peak
(called here ``Bragg peak").
The box size $L=256$ allows only a direct and fair determination of $g(x,N)$ for $x > 0.1$.
We have been forced to increase the box size to $L=512$ for smaller $x$ as may be seen
for an example with $x=0.1$ (dash-dotted line). 
As shown by the bold dashed line, the RPA formula is used to improve the estimation of $g(x,N)$
for small $x$.
\label{fig_Sq}
}
\end{figure}

\begin{figure}
\centering
\includegraphics[width=14cm]{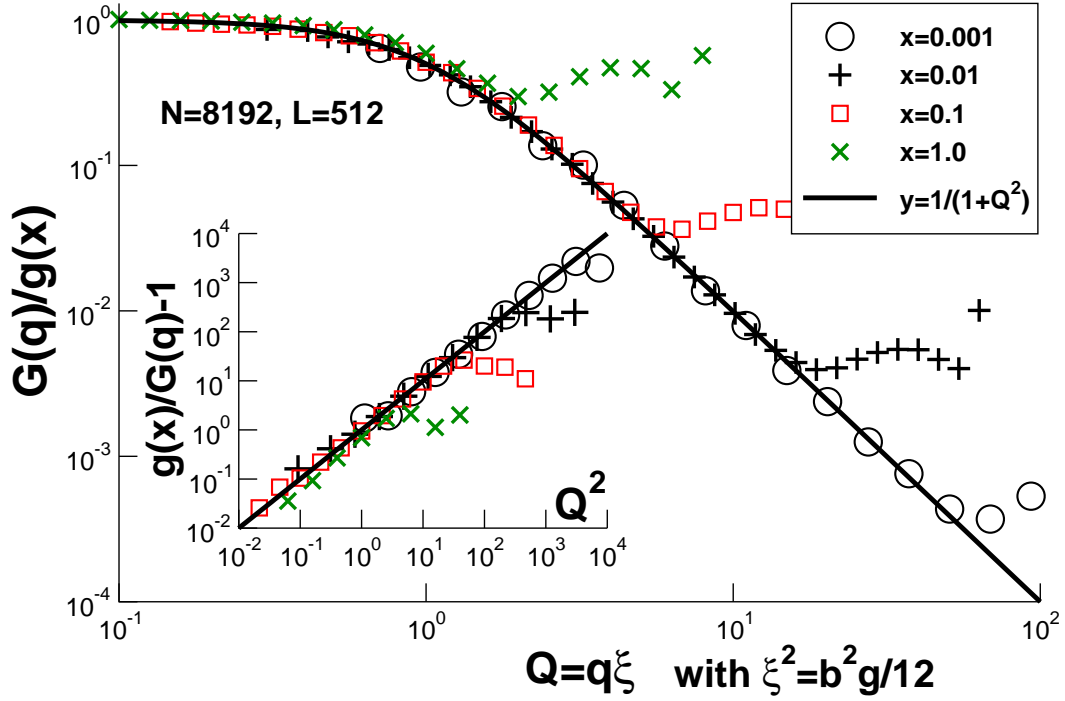}
\caption{Rescaled total structure factor $G(q)/g(x)$ 
as a function of the reduced wavevector $Q \equiv q\xi$ 
for chain length $N=8192$ and several $x \le 1$ as indicated.
The screening length $\xi$ of the thermal blob is obtained according to Eq.~(\ref{eq_xidef})
using the dimensionless compressibility $\geff(x)$ and the effective bond length $\be(x)$ from
Table~\ref{tab_epsasym}. The bold line compares the data with the approximated RPA,
Eq.~(\ref{eq_RPAapprox}). If replotted as indicated in the inset 
the data collapse on the bisection line. Deviations from the RPA formula 
become visible for larger $x$ as shown for $x=1$ (crosses).
\label{fig_Sqxi}
}
\end{figure}

\begin{figure}
\centering
\includegraphics[width=14cm]{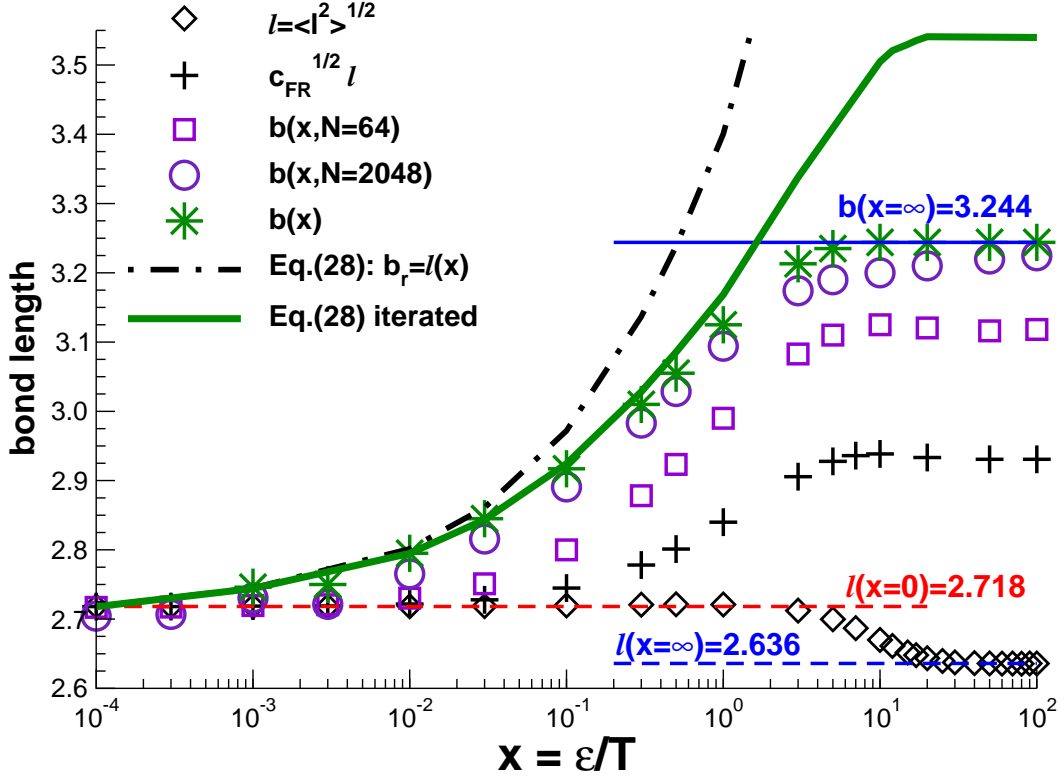}
\caption{The (effective) bond length as a function of the reduced overlap penalty $x=\overlap/T$.
The data for the root-mean-square bond length $l(x)$ and 
the effective bond length $\be(x)$ for asymptotically long chains
are listed in Tab.~\ref{tab_epsasym}.
%
%
The dash-dotted line indicates the effective bond length
as predicted by Eq.~(\ref{eq_bepredict}) assuming $\br = l(x)$ 
for the bond length of the reference chain.
The bold line shows the fix points obtained by iteration of Eq.~(\ref{eq_bepredict})
using as an input for the Ginzburg parameter the effective
bond length of the previous iteration step: $\be \rightarrow \br$.
See the main text for details. 
}
\label{fig_be}
\end{figure}

\begin{figure}
\centering
\includegraphics[width=14cm]{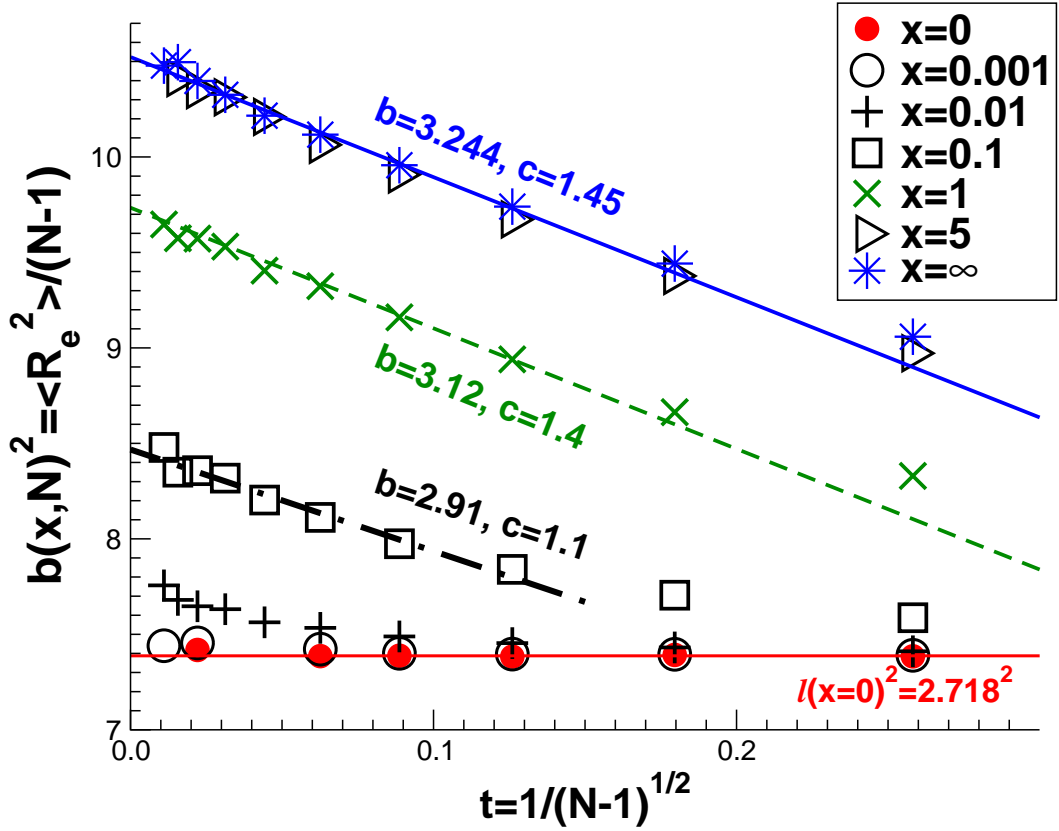}
\caption{Rescaled mean-square chain end-to-end distance 
$b(x,N)^2 \equiv \Rend^2(N)/(N-1)$
as a function of $t=1/\sqrt{N-1}$ for different $x$ as indicated.
The chains only remain Gaussian on all scales and all $N$ for extremely small $x$.
For $x \ge 0.1$ one observes $b(x,N)^2$ to decay linearly  
in agreement with Eq.~(\ref{eq_RNfit}).
This can be used for a simple two-parameter fit for $\be(x)$ as indicated for $x=0.1$, $1.0$ and $\infty$.
Note that the coefficient $c$ is slightly above unity as expected from Eq.~(19) of Ref.~\cite{WBM07}.
}
\label{fig_ReN}
\end{figure}

\begin{figure}
\centering
\includegraphics[width=14cm]{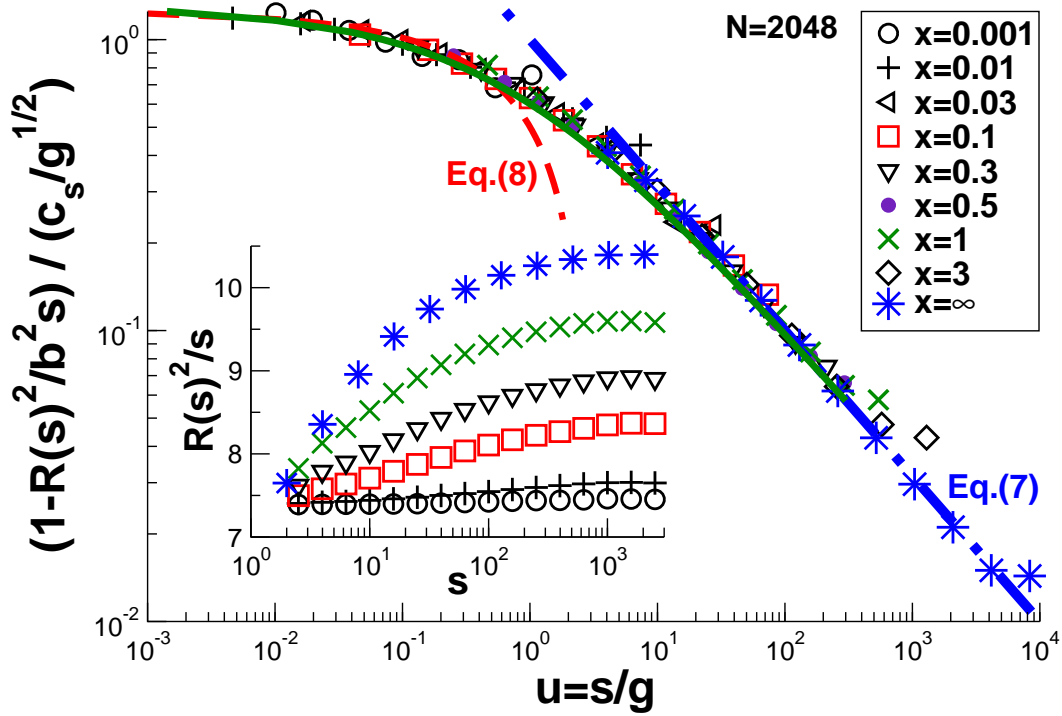}
\caption{Segment size $R(s)$ for overlap penalty $x$ as indicated for chain length $N=2048$.
Inset: $R(s)^2/s$ as a function of segment length $s$ increases monotonously 
approaching from below the asymptotic limit for large $s$, i.e. the chains are swollen.
Main figure: 
As suggested by Eq.~(\ref{eq_Rsgen}), the rescaled data 
$\left(1-R^2(s)/\be^2(x) s\right) / \left(\ce(x)/\geff(x)^{1/2}\right)$ 
is plotted as a function of the reduced arc-length $u=s/g$. 
The data collapse is successful for $1 \ll s \ll N$
which confirms the values $\geff(x)$ and $\be(x)$ for asymtotically long chains 
(Table~\ref{tab_epsasym}).  
The bold line shows the full prediction from Eq.~(\ref{eq_Rsgen}). 
We indicate the limiting behavior for small and large $u$ by the dashed and dash-dotted lines,
representing respectively Eq.~(\ref{eq_Rsfixm}) and Eq.~(\ref{eq_Rsmelt}).  
\label{fig_Rsscal}
}
\end{figure}

\begin{figure}
\centering
\includegraphics[width=14cm]{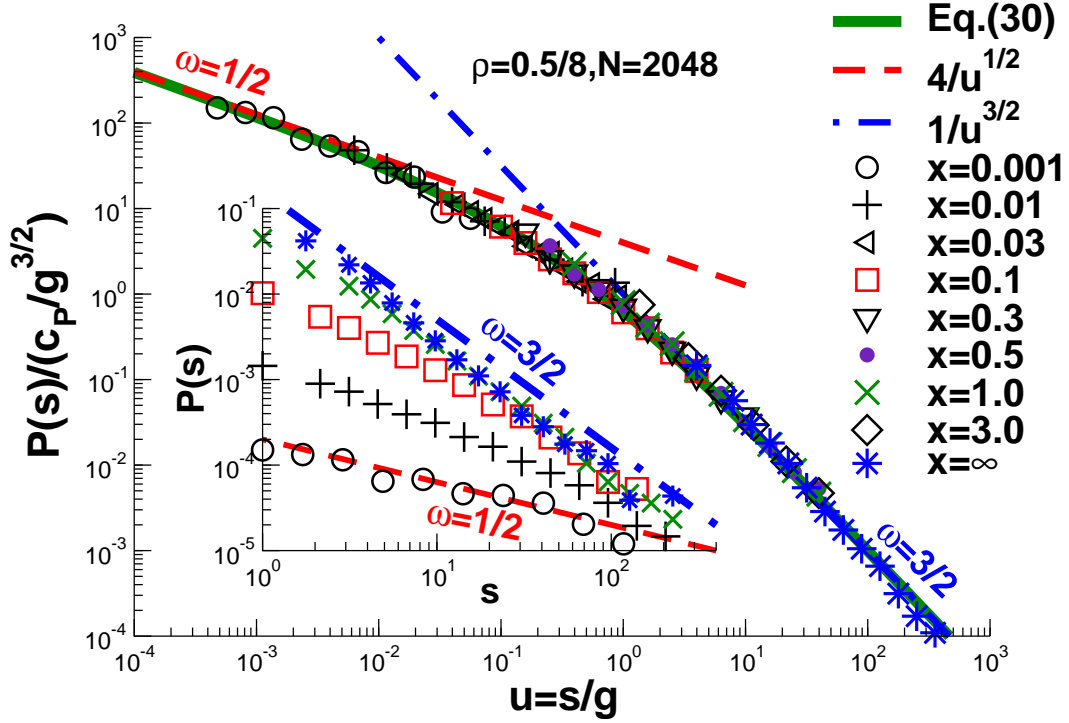}
\caption{Bond-bond correlation function $P(s)$ for different overlap penalties $x$ as indicated in the figure.
Inset: $P(s)$ as a function of segment length $s$ in log-log coordinates. 
The data approaches a power law behavior, $P(s) \sim 1/s^{\omega}$, 
with exponent $\omega =1/2$ for small $x$ (dashed line) and $\omega=3/2$ for $x \ge 1$ (dash-dotted line).
Main panel: Rescaled bond-bond correlation function $P(s)/\left[\cP(g)/g^{3/2}\right]$ plotted 
as a function of $u=s/g$ as suggested by Eq.~(\ref{eq_Psgen}). 
For large $u$, where an incompressible packing of thermal blobs is probed, all data collapse onto 
the dash-dotted line as predicted by Eq.~(\ref{eq_Psmelt}), i.e. $P(s)$ becomes independent 
of the compressibility $g$.
That this holds not only for the classical BFM with $x=\infty$ (stars) 
but also for finite $x$ is the central result of this study.
}
\label{fig_Ps}
\end{figure}

\paragraph*{Thermodynamic properties of a BFM version with finite overlap penalty.}
In this paper we have discussed a generalization of the standard bond-fluctuation model (BFM) 
where the monomers may overlap subject to a finite energy penalty $\overlap$ (Fig.~\ref{fig_BFMsketch}). 
This allows us to switch on systematically the excluded volume interaction between the monomers 
as suggested by perturbation theory \cite{DoiEdwardsBook} and to tune the density fluctuations 
of the solution at constant monomer density.
In this study we have focused on dense polymer melts containing flexible linear chains 
which are athermal apart from the finite overlap penalty. 
%
%
The central thermodynamic parameter characterizing these systems is the excess part of the dimensionless 
compressibility $g = T \kappaT \rho$ of the solution which has been obtained directly from the 
low-wavevector limit of the static structure factor (Figs.~\ref{fig_Sq} and \ref{fig_Sqxi}). 
Scanning the overlap penalty (or, equivalently, the temperature $T$ \cite{foot_x})
from $x = \overlap/T = \infty$ (no overlap) down to $x = 0.0001$
leads to a variation of $g(x)$ over four orders of magnitude (Fig.~\ref{fig_geps}).
This allows for a systematic study of the thermodynamic properties (Figs.~\ref{fig_Eeps}-\ref{fig_geps})
and the intrachain configurational statistics (Figs.~\ref{fig_be}-\ref{fig_Ps}). 
Particular attention has been paid to the thermodynamic properties of weakly interacting melts
($x \ll 1$). We have verified that our results are consistent with the free energy, Eq.~(\ref{eq_fThigh}), 
postulated in agreement with Edwards \cite{DoiEdwardsBook}. The main result
of this part of our study is that we have been able to demonstrate the density fluctuation 
contribution to the free energy induced by the chain connectivity from the scaling of the specific heat $\cV$ 
with respect to overlap penalty $x$ (Fig.~\ref{fig_cVeps}). 

\paragraph*{Intrachain conformational properties: Violation of Flory's ideality hypothesis.}
The broad variation of $g(x)$ puts us into a position to test the recently proposed Eq.~(\ref{eq_Rsgen})  
predicted by perturbation theory \cite{papRouse,papPs} describing the systematic swelling of chain segments
as function of the segment size $s$ and the compressibility $g(x)$. 
As outlined in the Introduction, this key relation suggests that the repulsive interactions between 
chain segments in the same chain are not fully screened at variance to Flory's ideality hypothesis for
polymers in dense melts \cite{DegennesBook}. 
The violation of the ideality hypothesis is demonstrated in Fig.~\ref{fig_Rsscal} for the 
mean-square segment size $R^2(s)$ and in Fig.~\ref{fig_Ps} for the bond-bond-correlation function $P(s)$. 
We show that data obtained for systems with very different compressibilities $g(x)$ can be superimposed on 
the predicted master curves if plotted as a function of the reduced arc-length $u=s/g$.
The scaling of $R(s)$ allows a precise determination of an important intrachain property, 
the effective bond length $\be(g)$ for asymptotically long chains (Fig.~\ref{fig_be}). These values 
compare well for $x < 1$ with the fix-point solution of the recursion relation, 
Eq.~(\ref{eq_bepredict}) \cite{DoiEdwardsBook,ME82,WBM07}.  
The bond-bond correlation function $P(s)$ being the second derivative of $R^2(s)$ with respect to $s$
allows an even more direct test of the predicted deviations. The reason is that the large Gaussian 
contribution $\be^2s$, which must be subtracted from $R^2(s)$ (see the vertical axis of Fig.~\ref{fig_Rsscal}), 
drops out due to the differentiation. In contrast to Flory's hypothesis, $P(s)$ does not vanish rapidly
on scales corresponding to the local persistence length (Fig.~\ref{fig_Ps}). In perfect agreement with 
theory \cite{foot_Ps}, the scaling plot shows two power law regimes characterized by exponents 
$\omega=1/2$ for small ($u \ll 1$) and $\omega=3/2$ for large ($u \gg 1$) reduced arc-length.
%
%

The central result of this study is that even for polymer melts with finite overlap penalty excluded volume interactions 
are not fully screened. If distances smaller than the thermal blob size are probed the chains are swollen according 
to the standard Fixman parameter expansion. More importantly, even on distances larger than the thermal blob size 
($s/g \gg 1$) 
deviations from ideal chain behavior are found. Interestingly, in this limit the explicit compressibility dependence
drops out and the relations established for incompressible melts \cite{WMBJOMMS04,WBM07} are recovered.
This shows that soft melts behave on large scales as incompressible packings of blobs.

\paragraph*{Outlook.}
Since the presented soft BFM is fully ergodic (in contrast to the classical BFM) 
and very efficient, it may be an interesting alternative to various popular coarse-grained
simulation approaches using soft effective interaction parameters 
\cite{Likos01,EM01,Guenza04,SCMF05,SCMF06}.
%
%
The presented model is part of a broader attempt to describe systematically the effects of
correlated density fluctuations in dense polymer systems, both for static
\cite{ANS96,ANS03,ANS05a,ANS05b} and dynamical \cite{ANS97,ANS98,MWBBL03} properties.
This also involves the comparison with (off-lattice) molecular dynamics simulation using a
standard bead-spring model which is discussed elsewhere 
\cite{MMP01,WBM07,papRouse,foot_persistence}.
An important unresolved question is for instance whether recently predicted long-range
{\em repulsive} forces of van der Waals type (``Anti-Casimir effect") \cite{ANS05a,ANS05b}
can be demonstrated numerically from specific non-analytic deviations from 
the RPA formula, Eq.~(\ref{eq_RPA}), at intermediate overlap strengths ($x \approx 1)$. 
In order to do this, we are currently improving the statistics of our data. 

In this paper we have discussed only {\em static} properties of the soft BFM.
Similar scaling behavior also has been obtained for the static Rouse mode correlation function
which displays systematic deviations from the scaling expected for ideal chains \cite{papRouse}. 
We currently are working out how these deviations 
may influence the {\em dynamics} for polymer chains {\em without} topological constraints.
(These constraints can be switched off even for $x=\infty$ by using the ``L26" local moves
described in Sec.~\ref{sec_algo}.) Conceptually important issues can be addressed if the 
(artifical) slithering-snake dynamics is analysed and compared to predictions of the 
``activated-reptation dynamics" hypothesis suggested by Semenov \cite{ANS97,ANS98} 
for real, although extremely long polymer chains. If no overlap is allowed,
the slithering-snake dynamics is known to show anomalous curvilinear diffusion and 
correlated motion of neighboring snakes \cite{MWBBL03}. 
Since for $x=\infty$ the lattice might influence the results, it is important to verify
if qualitative similar behavior is also found for soft BFM melts with thermal blobs much larger than
the local monomer scale and how the anomalous curvilinear diffusion changes with compressibility.

\begin{acknowledgments}
We thank H.~Meyer and A.N.~Semenov (both ICS, Strasbourg, France), S.P.~Obukhov (Gainesville, Florida) 
and M. M\"uller (G\"ottingen) for helpful discussions.
A generous grant of computer time by the IDRIS (Orsay) is also gratefully acknowledged.
We are indebted to the Universit\'e de Strasbourg and the ESF STIPOMAT programme for financial support.
J.B. acknowledges financial support by the IUF.
\end{acknowledgments}



\begin{thebibliography}{79}
\bibitem[{foo({\natexlab{a}})}]{foot_asymlimit}
\bibinfo{note}{The dimensionless compressibility $g$ and the effective bond
  length $b$ depend in general on both the overlap strength $x=\overlap/T$ and
  the chain length $N$. We refer to the limiting values $\geff(x) =
  \lim_{N\to\infty} g(x,N)$ and $\be(x) = \lim_{N\to\infty} b(x,N)$ of
  asymptotically long chains if not stated otherwise. The chain length
  dependence of the compressibility is due to the translational entropy of the
  chains [Eq.~(\ref{eq_gNeffect})]. The much stronger $N$-effects observed for
  the effective bond length are due to swelling of the chains as described by
  Eq.~(\ref{eq_Rsgen}) and discussed in Sec.~\ref{sub_chainsize}.}
\bibitem[{foo({\natexlab{b}})}]{foot_Gq}
\bibinfo{note}{For the computational definition of $G(q)$ see
  Sec.~\ref{sub_compress}. Note that our normalization of the structure factor
  is consistent with Eq.~(5.39) of \cite{DoiEdwardsBook}, i.e. $G(q)$ is
  dimensionless. Quite generally, we present in this paper properties
  normalized by the total number of monomers $\nmon$, i.e. essentially by the
  degree of freedom of the system, rather than by its volume $V$.}
\bibitem[{foo({\natexlab{c}})}]{foot_veff}
\bibinfo{note}{Following \cite{ANS05a,BJSOBW07,WBM07} the ``bulk modulus"
  $\veff \equiv 1/g\rho$ is sometimes called ``effective excluded volume" which
  refers to the fact that convergence and self-consistency of the perturbation
  theory \cite{DoiEdwardsBook} suggests to substitute (renormalize) the bare
  excluded volume $\vone$ of the monomers by the explicitly measured $\veff$
  summing also over higher virial contributions, not only pair interactions.
  However, since higher virial contributions become negligible for weak
  excluded volume interactions one expects that to leading order $v(x) \approx
  \veff(x)$ for $x \ll 1$. This will be crosschecked in Sec.~\ref{sub_compress}
  (Fig.~\ref{fig_geps}). Since the perturbation calculation summarized by
  Eq.~(\ref{eq_fThigh}) is anyway restricted to the same $x$-regime
  [Eq.~(\ref{eq_Ginzburg})] we do not use $\veff$ in the main text.}
\bibitem[{foo({\natexlab{d}})}]{foot_x}
\bibinfo{note}{Since the overlap penalty is the only energy scale in this study
  one may either vary the overlap parameter $\overlap$ or the temperature $T$.
  Since the presentation of thermodynamic properties (especially
  Figs.~\ref{fig_Eeps}-\ref{fig_mueps}) becomes, however, slightly simpler if
  $T$ is the control parameter we fix arbitrarily $\overlap=1$. The inverse
  temperature $\beta=1/T$ and the dimensionless overlap strength $x=\overlap/T$
  are thus numerically equal. We keep both notations for dimensional reasons
  and for future generalization to models with more than one energy scale.}
\bibitem[{foo({\natexlab{f}})}]{foot_connectivity}
\bibinfo{note}{Interestingly, according to Edwards \cite{DoiEdwardsBook} the
  chain connectivity, i.e. the presence of attractive forces between bonded
  monomers, does not change the excluded volume $v(x)$ --- as one would expect
  naively --- but rather gives rise to an additional term scaling differently
  with density.}
\bibitem[{foo({\natexlab{g}})}]{foot_Ginzburg}
\bibinfo{note}{The Ginzburg parameter is defined here in terms of the measured
  compressibility $g(x)$ and the measured effective bond length $\be(x)$.
  Eq.~(\ref{eq_Ginzburg}) is consistent with Eq.~(5.46) of
  Ref.~\cite{DoiEdwardsBook}.}
\bibitem[{foo({\natexlab{h}})}]{foot_fourier}
\bibinfo{note}{Since the intrachain form factor is essentially the Fourier
  transform of the segmental size distribution \cite{WBM07,BJSOBW07}, $R(s)$ is
  related to an experimentally relevant quantity.}
\bibitem[{foo({\natexlab{j}})}]{foot_attraction}
\bibinfo{note}{Attractive interactions similar to the ones used in
  \cite{MarcusT3D95,Blends99,Blends03} may be easily added to the Potts spin
  formulation of the soft BFM. The simulation of polymer blends requires
  additional Potts spin lattices interacting as indicated in
  Eq.~(\ref{eq_ghostinter}) for the two lattices used to obtain the chemical
  potential in Sec.~\ref{sub_mu}.}
\bibitem[{foo({\natexlab{k}})}]{foot_muN}
\bibinfo{note}{Small corrections to $\geff(x) \sim N^0$ and $\muex \sim N$ may
  arise due to the long-range correlations which lead to the swelling described
  by Eq.~(\ref{eq_Rsgen}). This effect is, however, too weak to be visible in
  our data. It can be demonstrated from the number distribution of equilibrium
  polymers \cite{WMC98b,WBCHCKM07} which differs slightly from the exponential
  behavior generally admitted \cite{ANS03,BJSOBW07}.}
\bibitem[{foo({\natexlab{l}})}]{foot_Ps}
\bibinfo{note}{Our theoretical result, Eq.~(\ref{eq_Psgen}), has already been
  applied successfully to the bond-bond correlation function obtained from
  single-chain-in-mean-field simulation \cite{SCMF06}. Over the range of
  arc-lengths $s$ studied in Ref.~\cite{SCMF06} ($0.04 \le u =2/g < 4$)
  Eq.~(\ref{eq_Psgen}) accounts for the shape of $P(s)$, but quantitatively
  lies slightly above the simulation data, perhaps because the effective bond
  length $\be$, which is determined from $\Rend(N)$ [rather than from $R(s)$ as
  suggested in Sec.~\ref{sub_chainsize}], could be refined.}
\end{thebibliography}

\newpage

\begin{table}[t]
\begin{tabular}{|c||c|c|c|c|c|c|c|c|c|}
\hline
$x$   & $e$    & $\cV$   &$\muex/T$& $g$     \\ \hline
0.0001& 0.2499 & 1.14E-09& 5.0E-05 &  $\approx 1$ \\
0.001 & 0.2499 & 1.08E-07& 5.0E-04 &  $\approx 1$ \\
0.01  & 0.2489 & 1.05E-05& 0.0049  &  $\approx 1$ \\
0.1   & 0.2400 & 0.00093 & 0.049   &  0.95\\
0.3   & 0.2228 & 0.00709 & 0.14    &  0.88\\
1.0   & 0.1799 & 0.04854 & 0.43    &  0.69\\
3.0   & 0.1147 & 0.21538 & 1.04    &  0.47\\
10    & 0.0321 & 0.57795 & 2.15    &  0.25\\
20    & 0.0223 & 0.36671 & 2.55    &  0.20\\
30    & 0.0012 & 0.16318 & 2.62    &  0.20\\
50    & 8.9E-05& 0.02874 & 2.63    &  0.20\\
100   & 2.2E-07& 0.00039 & 2.635   &  0.20\\
$\infty$& 0    & 0       & 2.635   &  0.20 \\ 
\hline
\end{tabular}
\vspace*{0.5cm}
\caption[]{Various properties for soft BFM beads ($N=1$) at monomer number density $\rho = 0.5/8$
(corresponding to a volume fraction $0.5$) and linear box size $L=256$ 
as a function of the reduced overlap strength $x=\overlap/T$.
The limit $x=\infty$ corresponds to the classical BFM without monomer overlap,
the limit $x=0$ to non-interacting monomers. Indicated are
the mean energy per bead $e$, the specific heat $\cV$ per bead,
the excess part of the chemical potential $\muex/T$,
and the dimensionless compressibility $g(x,N=1)$.
Within statistical accuracy we obtain below $x \approx 0.1$ the ideal gas compressibility, $g \approx 1$, 
and above $x \approx 10$ the compressibility for a melt without monomer overlap.
\label{tab_beads}}
\end{table}

\newpage
\begin{table}[t]
\begin{tabular}{|c||c|c|c|c|c|c|c|c|c|c|c|}
\hline
$x$     & $e$    &$\eself$& $\cV$  &$\muex/(TN)$& $g$ & $l$ &$\be$ & $\la \theta \ra$ & $\la\cos(\theta)\ra$ & $\ce$ & $\Gi$ \\ \hline
0       & 0      & 0      &  0     & 0        &$\infty$ &2.718 & 2.72  & $90^{\circ}$     & 0                    & -     & 0    \\
0.0001  & 0.42   & 0.18   & 2.5E-07& 4.9E-05  & 20094   &2.718 & 2.75  & $90^{\circ}$     & 0                    & 0.68  & 0    \\
0.001   & 0.42   & 0.18   & 8.8E-06& 4.9E-04  & 2029    &2.718 & 2.75  & $89.99^{\circ}$  & 1.7E-04              & 0.68  & 0.017\\
0.01    & 0.39   & 0.17   & 2.2E-04& 4.5E-03  & 209     &2.718 & 2.80  & $89.9^{\circ}$   & 1.1E-03              & 0.65  & 0.052\\
0.1     & 0.32   & 0.15   & 4.5E-03& 0.05     & 22      &2.719 & 2.92  & $89.4^{\circ}$   & 9.2E-03              & 0.57  & 0.14 \\
0.3     & 0.26   & 0.12   & 0.015  & 0.1      & 7.1     &2.720 & 3.01  & $88.5^{\circ}$   & 0.021                & 0.52  & 0.22 \\
1       & 0.18   & 0.08   & 0.06   & 0.4      & 2.4     &2.721 & 3.13  & $86.9^{\circ}$   & 0.043                & 0.46  & 0.34 \\
3       & 0.11   & 0.05   & 0.3    & 0.9      & 0.85    &2.721 & 3.21  & $84.9^{\circ}$   & 0.069                & 0.42  & 0.52 \\
10      & 0.03   & 0.01   & 0.5    & 1.8      & 0.32    &2.670 & 3.24  & $82.9^{\circ}$   & 0.096                & 0.41  & 0.83 \\
20      & 0.004  & 0.002  & 0.26   & 2.0      & 0.25    &2.643 & 3.24  & $82.4^{\circ}$   & 0.104                & 0.41  & 0.94 \\
30      & 9.7E-04& 4.0E-04& 0.11   & 2.0      & 0.25    &2.638 & 3.24  & $82.3^{\circ}$   & 0.105                & 0.41  & 0.94 \\
50      & 7.1E-05& 2.9E-05& 0.019  & 2.1      & 0.25    &2.636 & 3.24  & $82.2^{\circ}$   & 0.106                & 0.41  & 0.94 \\
100     & 2.2E-07& -      & 4.5E-04& 2.1      & 0.25    &2.636 & 3.24  & $82.2^{\circ}$   & 0.106                & 0.41  & 0.94 \\
$\infty$& 0      & 0      & 0      & 2.1      & 0.25    &2.636 & 3.24  & $82.2^{\circ}$   & 0.106                & 0.41  & 0.94 \\
\hline
\end{tabular}
\vspace*{0.5cm}
\caption[]{Various properties for asymptotically long BFM chains at number density $\rho = 0.5/8$
as a function of $x=\overlap/T$. Apart from the properties already presented in
Table~\ref{tab_beads} for beads we indicate here
the intrachain self-energy $\eself$,
the root-mean-square bond length $l = \la \bm{l}_n^2 \ra^{1/2}$,
the effective bond length $\be$,
the mean angle $\la \theta \ra$ and
the mean cosine $\la \cos(\theta) \ra = \la \bm{e}_n \cdot \bm{e}_{n+1} \ra$ of 
two subsequent bonds with $\bm{e}_n=\bm{l}_n/|\bm{l}_n|$ being the normalized bond vector, 
the swelling coefficient $\ce \equiv \sqrt{24/\pi^3}/\rho \be^3$,
and the Ginzburg parameter $\Gi = 1/\sqrt{g}\be^3\rho$.
The excess part of the chemical potential of a chain is given in units of the chain length $N$ (column 4).
The effective bond length $\be(x)$ has been obtained using an extrapolation scheme 
implied by Eq.~(\ref{eq_Rsmelt}) and discussed in Sec.~\ref{sub_chainsize}. 
%
%
\label{tab_epsasym}}
\end{table}

\newpage
\begin{table}[t]
\begin{tabular}{|c||c|c|c|c|c|c|}
\hline
$N$   & $e$    & $\cV$  &$\muex/(TN)$&$g$& $l$  &$b$\\ \hline
1     & 0.1799 & 0.0485 & 0.43     & 0.69& -    & -    \\
4     & 0.1767 & 0.0482 & 0.40     & 1.5 & 2.717& 2.77 \\
16    & 0.1811 & 0.0691 & 0.37     & 2.0 & 2.720& 2.89 \\
64    & 0.1819 & 0.0562 & 0.35     & 2.3 & 2.721& 2.99 \\
256   & 0.1820 & 0.0877 & 0.35     & 2.4 & 2.721& 3.05 \\
1024  & 0.1820 & 0.0871 & 0.34     & 2.4 & 2.721& 3.08 \\
2048  & 0.1820 & 0.0639 & 0.34     & 2.4 & 2.721& 3.09 \\
4096  & 0.1820 & 0.0697 & 0.34     & 2.4 & 2.721& 3.10 \\
8192  & 0.1820 & 0.0795 & 0.34     & 2.4 & 2.721& 3.11 \\
\hline
\end{tabular}
\vspace*{0.5cm}
\caption[]{Various properties for BFM melts of number density $\rho = 0.5/8$
at overlap strength $x=\overlap/T=1$ as a function of chain length $N$.
For small chains the overlap energy $e$ 
and its fluctuation $\cV$ increase slightly while the chemical potential per bead decreases. 
The compressibility $g(x=1,N)$ becomes chain length independent for $N > 64$.
The chain length dependence visible for small $N$ is described by Eq.~(\ref{eq_gNeffect}),
i.e. the data is consistent with an excess compressibility $\gex(x) \approx 2.4 N^0$ for all $N$.
The last column indicates the rescaled end-to-end distance $b(x=1,N) \equiv \Rend(N)/(N-1)^{1/2}$
which approaches the effective bond length $\be(x) \approx 3.13$ of asymptotically long chains 
monotonously from below, just as for classical BFM melts \cite{WBM07}.
Interestingly, $b(x=1,N)$ has not yet reached the asymptotic limit $\be(x)$ even for $N=8192$
albeit all other quantities indicated can be regarded (within statistical accuracy) 
as independent of chain length above $N\approx 256$.
\label{tab_eps1.0}}
\end{table}

\newpage
\clearpage

\newpage

\newpage

\newpage

\newpage

\newpage

\newpage

\newpage

\newpage
\clearpage

\newpage

\newpage
\clearpage

\newpage
\end{document}